# Slow, Nanometer Light Confinement Observed in Atomically Thin $TaS_2$

Hue T.B. Do[1,2,3#], Meng Zhao[2#*], Pengfei Li[4], Yu Wei Soh[2], Jagadesh Rangaraj[1,2], Bingyan Liu[5], Tianyu Jiang[2,5], Xinyue Zhang[1,2], Jiong Lu[4], Peng Song[5], Jinghua Teng[2], Michel Bosman[1,2*]

[1]Department of Materials Science and Engineering, National University of Singapore, 9 Engineering Drive 1, 117575, Singapore.

[2]Institute of Materials Research and Engineering (IMRE), Agency for Science, Technology and Research (A*STAR), 2 Fusionopolis Way, 138634, Singapore.

[3]NUS Graduate School - Integrative Sciences and Engineering Programme (ISEP), National University of Singapore, 21 Lower Kent Ridge Road, 119077, Singapore.

[4]Department of Chemistry, National University of Singapore, 3 Science Drive 3, 117552, Singapore.

[5]School of Electrical and Electronic Engineering, Nanyang Technological University, 50 Nanyang Avenue, 639798, Singapore.

*zhaom@imre.a-star.edu.sg; *msemb@nus.edu.sg

[#] equal contributions

## Abstract

Extreme light confinement down to the atomic scale has been theoretically predicted for ultrathin, Ta-based transition metal dichalcogenides (TMDs). In this work, we experimentally demonstrate in 2H-$TaS_2$ monolayers and bilayers a lateral confinement ratio up to 300 at large wave vectors of q = 0.15 $Å^{-1}$, and slow light behaviour with a group velocity ~$10^{-4}c$. Quantitative momentum-resolved electron energy loss spectroscopy (q-EELS) with a momentum resolution of 0.0056 $Å^{-1}$ was used as a platform for the nanoscale optical measurements. With it, momentum-dispersed, two-dimensional (2D) plasmon resonances were experimentally observed, showing a transition from 2D to 3D Coulomb interaction in the high-momentum regime, equivalent to light confinement volumes of 1-2 $nm^3$. Remarkably, the resonant modes do not enter the electron-hole continuum, predicting even further enhanced optical field confinements for this material at cryogenic temperatures.

# Introduction

The ultimate confinement of guided light takes place in atomically-thin films. Metallic monolayers in particular are predicted to provide a new platform in which light couples to spatially confined plasmons—collectively oscillating free electrons. This ability to focus light energy to sub-nanometer length scales makes metallic monolayers one of the prime candidates for enhanced light-matter interaction[1], deep sub-diffraction imaging and spectroscopy[2,3], and future nanoscale optical computing[4,5].

In-plane confinement is characterised by the ratio $\lambda_0/\lambda_p$, where $\lambda_0$ is the light wavelength in free space and $\lambda_p$ is the plasmon polariton wavelength in the material. This ratio is close to 200 for Dirac plasmons in graphene[4]. However, at large wave vector (small $\lambda_p$), graphene plasmons enter the electron-hole continuum where the plasmon provides sufficient momentum and energy to facilitate interband or intraband transitions, resulting in severe plasmon damping[6]. This damping mechanism also applies to other 2D electron gas systems such as Si surface states and monolayer Ag on Si[6,7], but is theoretically predicted to be absent in 2H-$TaS_2$.

Metallic Group V transition metal dichalcogenides (TMDs) such as 2H-$TaS_2$, $TaSe_2$, and $NbSe_2$ are a particularly promising class of 2D materials for light confinement. These materials have an isolated conduction band, allowing plasmons to disperse up to a much higher wave vector before entering the lossy electron-hole continuum[8]. Recently, reliable first-principles calculations predicted a flat dispersion relation in 2H-$TaS_2$ monolayers up to a large wave vector $q = 2\pi/\lambda_p$ of 0.4 Å$^{-1}$ in the near-infrared frequency range[9]. The predicted flat dispersion relation indicates ultra-slow light behaviour, resulting in a high confinement ratio and giant field enhancement.

In this work, we use monochromated momentum-resolved electron energy loss spectroscopy (q-EELS) to experimentally demonstrate the existence of highly confined plasmons at room temperature in 2D 2H-$TaS_2$ monolayers and bilayers. We observe the first flat plasmon dispersions in 2H-$TaS_2$, both in monolayers and bilayers, at least up to 0.15 Å$^{-1}$ as predicted theoretically[9]. The corollary of this observation of a flat plasmon dispersion, is that it provides evidence for the existence of ultra-slow plasmon polariton waves in atomically thin metals. In contrast to previous q-EELS studies on bulk metallic TMDs[10,11], our work is performed in the true monolayer limit, enabling the direct experimental observation of nonlocal interband screening—proposed as the key mechanism responsible for the peculiar flat plasmon dispersion, and potentially universal across all 2D metallic TMD systems.

By their nature, highly dispersed plasmons have a large momentum-mismatch with light, preventing them to be excited directly with far-field light illumination. To excite and measure these plasmons, several methods can provide the missing momentum. For example, the material can be patterned in an array and measured with Fourier transform infrared spectroscopy (FTIR)[12]. Alternatively, the plasmons can be excited in the near-field, as in scattering-type scanning near-field optical microscopy (s-SNOM)[13–15] or with a focused electron beam. However, for FTIR and s-SNOM, the accessible momentum range is limited to 0.01 Å$^{-1}$, which makes it difficult to observe details on the dispersion, which develops at values well beyond 0.01 Å$^{-1}$. A fast-moving electron beam on the other hand, can provide momentum all across the Brillouin zone up to > 1 Å$^{-1}$ [16,17], making it useful for quantitative plasmon dispersion measurements. In previous work, the scanning transmission electron microscope (STEM) was used for q-EELS measurements on ~100 nm thick 3D bulk metallic TMD samples of 2H-$TaS_2$, $TaSe_2$, $NbS_2$ and $NbSe_2$, showing a remarkable negative plasmon dispersion at large wave vectors[10,11]. This effect was proposed to originate from the detailed structure of the narrow d-band near the Fermi level[18–20]. However, measurements on atomically-thin metallic TMDs have not yet been demonstrated so far, preventing experimental evidence for the predicted local field screening, slow light, and its confinement in the monolayer limit.

Fig. 1a schematically shows the under-focused electron beam projecting the reciprocal lattice of the free-hanging $TaS_2$ at the entrance of the EELS spectrometer, an acquisition method that has been shown to deliver high momentum-resolution in the STEM[21–23]. With this experimental set-up, individual EELS spectra with small collection angles are acquired at different positions in momentum space with typical energy resolutions of 50 meV and momentum-resolutions of 0.0056 $\text{Å}^{-1}$. It provides the conditions to explore whether plasmons indeed avoid the electron-hole continuum and encounter the predicted conditions for low plasmon damping.

High-quality $TaS_2$ flakes were obtained by electrochemical exfoliation[24,25], a method that produces large-area, ultrathin flakes with high yield while preserving the crystal structure. X-ray diffraction shown in Supplementary Information (SI) Fig. S1 confirms that the original $TaS_2$ crystal used in this work is in the 2H phase, with high crystallinity. After exfoliation, the solution was drop-cast on $SiO_2$/Si substrates following washing and liquid cascade centrifugation. The optical image in Figure 1b demonstrates that the as-exfoliated flakes are uniform monolayers with average lateral sizes exceeding 100 µm. Selected area electron diffraction (SAED) measurements, as shown in Fig. 2a and Fig. S4 confirm the high crystallinity of the as-exfoliated $TaS_2$ flakes. Atomic resolution images of monolayer and bilayer samples are provided in Fig. S2, showing that the atomic structure is conserved in the exfoliated films. The quality of the exfoliated $TaS_2$ monolayers was further examined by superconductivity measurements, which show a transition temperature of 3 K, close to the previously reported 3.4 K for high-quality, mechanically exfoliated $TaS_2$ monolayers[26]. The two peaks in the measured Raman spectra in Fig. 1c at 301 $cm^{-1}$ and 396 $cm^{-1}$ correspond to the $A_{1g}$ and $E_{2g}^{1}$ phonon modes, confirming the 2H-phase[27,28]. The soft mode at 183 $cm^{-1}$ can be attributed to the two-phonon scattering at room temperature[29].

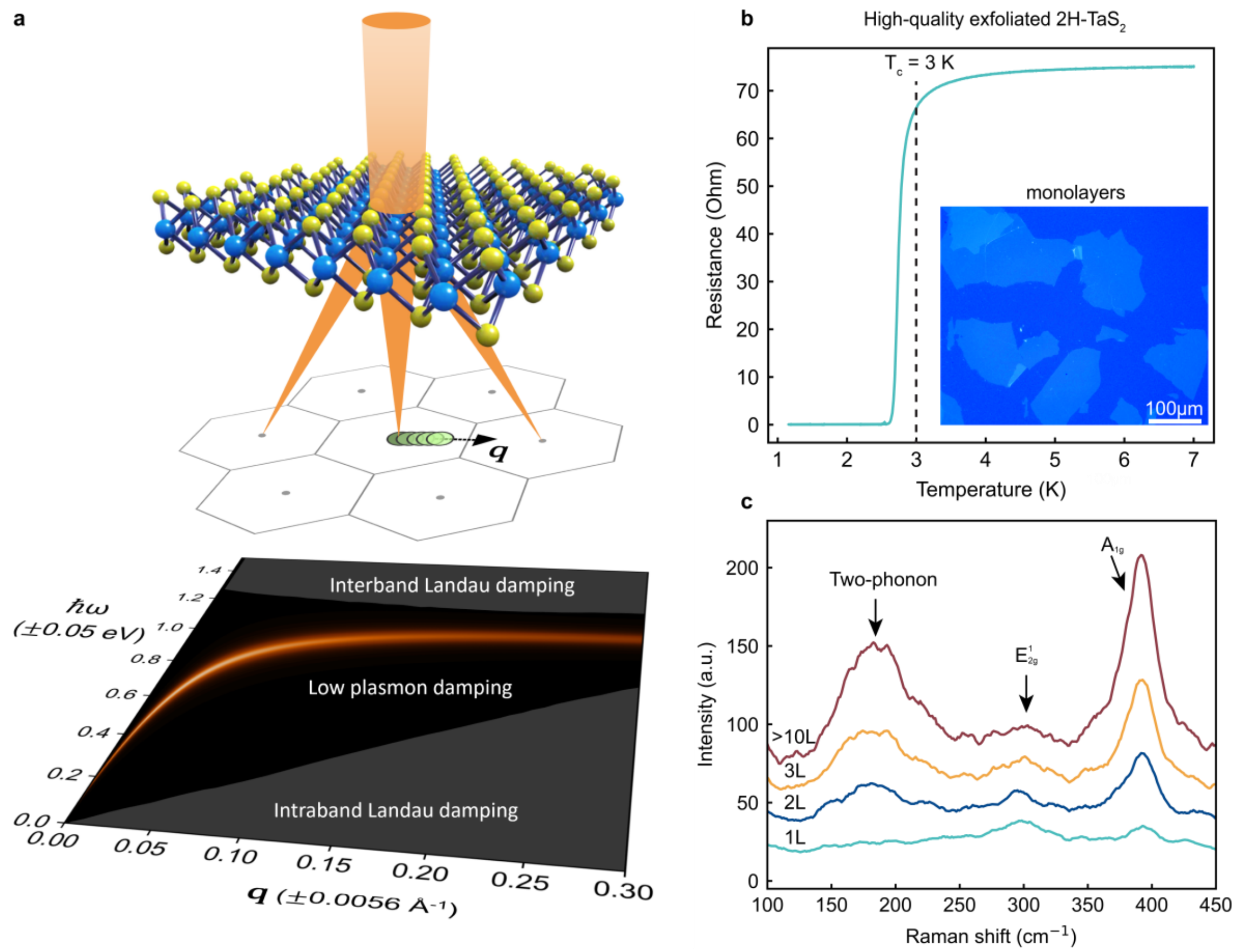

**Fig. 1 | Experimental set-up for q-EELS on 2H-$TaS_2$ metallic monolayers. a** Schematic of the experimental set-up, using a near-parallel, under-focused electron beam (orange) in a STEM for monochromated EELS with very high momentum resolution (0.0056 Å$^{-1}$) in reciprocal space. 2D plasmons are predicted to disperse outside the electron-hole continuum, indicated by the dark region between the interband and intraband Landau damping regions[9]. **b** Bright field light microscope image of 2H-$TaS_2$ monolayers after electrochemical exfoliation, and experimental demonstration of superconductivity, demonstrating the high quality of the used sample material. **c**, Layer-dependent Raman spectroscopy of the 2H-$TaS_2$.

# Results and discussion

Three momentum-dispersed parameters were extracted from our measurements: the plasmon peak position, its intensity, and its width. The peak position is used to quantify the confinement ratio in the slow-light regime as well as the interband screening that is responsible for the flat dispersion relation. From the plasmon peak intensity, we demonstrate the transition from 2D to 3D Coulomb interaction at large wave vectors. Finally, the plasmon peak width measurements show the dominance of phonon-assisted Landau damping for these plasmons at room temperature.

Fig. 2a presents a series of momentum-dispersed EELS spectra along the $\Gamma - M$ direction in reciprocal space, with the individual spectra displaced in the y-direction proportional to their momentum. The green disks in the inset of Fig. 2a represent the collection angles ($\Delta q_\beta$) of the EELS spectra with the corresponding colour. The EELS spectrum collected in each green disk is the probability density $\frac{d^2P}{d\Omega dE}(\omega, q)$ of the energy loss $E = \hbar\omega$ and the in-plane momentum transfer $\hbar\vec{q}$ in a solid angle $\Omega$. The EELS probability density firstly depends on the Coulomb interaction between the fast electron beam and the plasmons, described by the EELS prefactor $I_{kin}(\omega, q)$, as derived in SI Section 3. Secondly, it depends on the loss function $\mathcal{L}(\omega, q) = Im(\varepsilon_{2D}^{-1}(\omega, q))$ of the sample material, which is described by its dielectric function:

$$\frac{d^2P}{d\Omega dE}(\omega, q) = I_{kin}(\omega, q) \times Im(\varepsilon_{2D}^{-1}(\omega, q)) \tag{1}$$

For the case of a 2D electron gas with vanishing thickness $d$ ($qd \ll 1$) in the non-relativistic limit with negligible radiation loss ($\omega \ll cq$), the EELS prefactor $I_{kin}(q)$ scales with $q^{-3}$ (Equation S14), agreeing with previous formalisms for the case of graphene[30,31].

As suggested by da Jornada et al.[9], the optical response of $TaS_2$ monolayers can be modelled as the response of an ideal 2D electron gas embedded in a dielectric environment that accounts for a substrate and for interband screening $\varepsilon_{inter}(q)$. A similar approach was implemented to model 2H-$TaSe_2$ thin films (>10 nm) by Song et al.[12]. Using this approach in SI Section 2, we show that the loss function $Im(\varepsilon_{2D}^{-1}(\omega, q))$ for free-hanging 2H-$TaS_2$ monolayers and bilayers can be written as a Lorentz-Drude function (Equation S16), characterised by the plasmon angular frequency $\omega_p(q)$ and damping rate $\Gamma(q)$. For each EELS spectrum, we remove the quasi-elastic background (Fig. S6) and fit the plasmon peak with a Lorentz-Drude function. In this way, we obtain for each spectrum a specific set of fitting parameters:

- The plasmon peak position:

$$\omega_p(q) = \sqrt{\frac{\mathcal{D}}{2\pi\varepsilon_0}\frac{q}{\varepsilon_{inter}(q)}}, \tag{2}$$

with the Drude weight of the 2D electron gas $\mathcal{D} = \pi e^2 \left[\frac{n}{m}\right]_{eff}(q)$.

- The peak intensity:

$$A(q) = I_{kin}(q) \times \frac{1}{\varepsilon_{inter}(q)}, \tag{3}$$

- And the peak width: $\Gamma(q)$

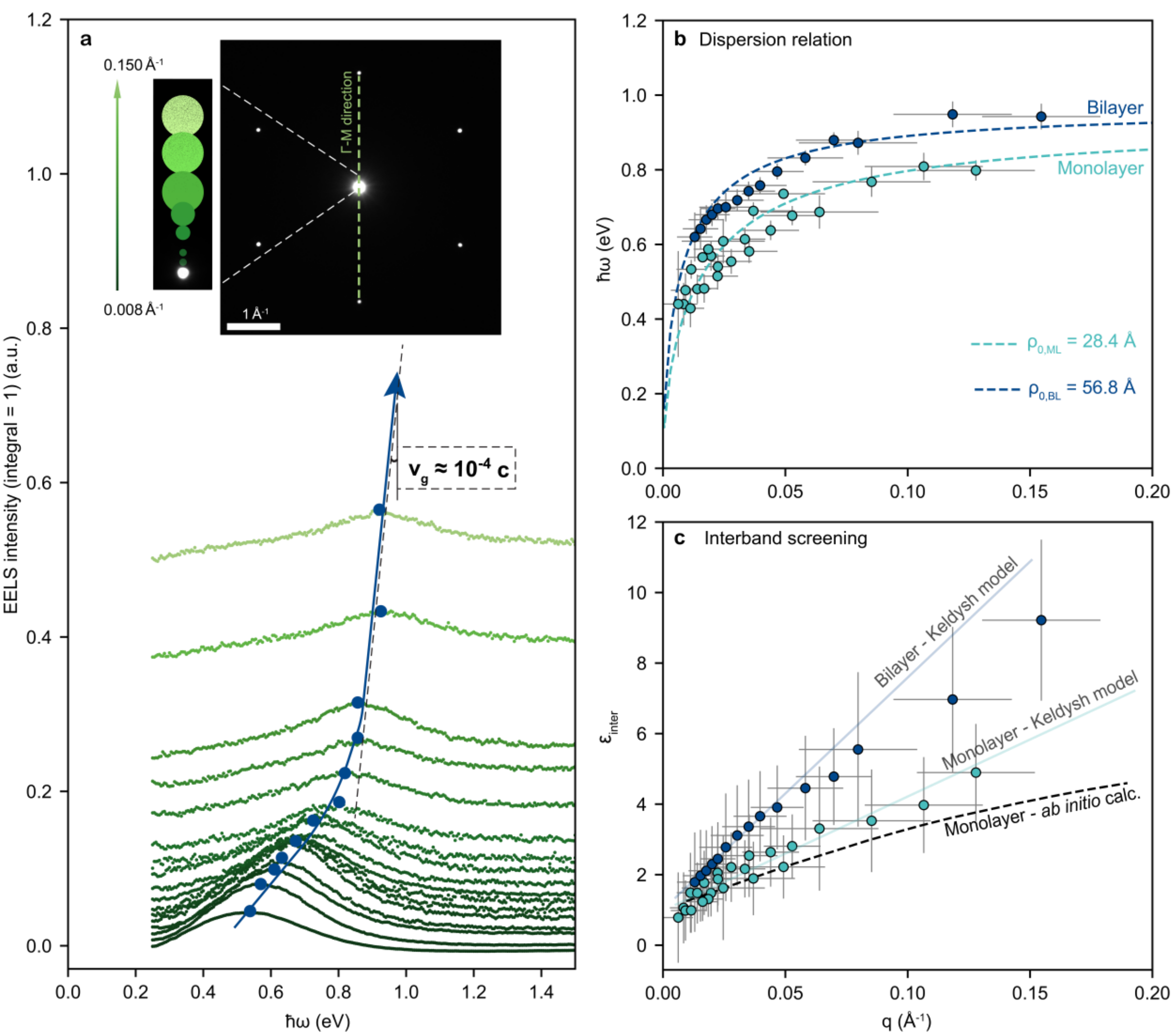

**Fig. 2 | q-EELS measurements of monolayer and bilayer 2H-$TaS_2$ along the Γ-M direction. a** Bilayer spectra, vertically shifted based on the corresponding q-value set in reciprocal space, as indicated by the coloured disks in the inset. 2D plasmon peaks are observed and fitted with a Drude-Lorentz function to extract the q-dependent plasmon energy, forming a dispersion relation. The black dashed line in Fig. 2a serves as a visual guide to estimate the group velocity from the slope of the fitted plasmon dispersion relation. The lowest group velocity observed in the measurement is estimated to be $\sim 10^{-4}c$ at high momentum. **b** Fitting of the measured dispersion relation of the monolayer (cyan) and bilayer (blue) $TaS_2$ with the model based on Equation (2). **c** Interband screening component $\varepsilon_{inter}$ of the dielectric function calculated with screening lengths $\rho_{0,ML} = 28.4$ Å for the monolayer and $\rho_{0,BL} = 56.8$ Å for the bilayer. The solid cyan and blue lines represent the linear q-dependence expected from the Keldysh model of Equation (4), serving as a visual guide.

## Slow light behaviour and strong confinement

In this section and the next, the measured plasmon frequency, i.e. the peak position, is analysed, to quantify the group velocity and strong confinement of plasmons in $TaS_2$ mono- and bilayers. The scatter plots in Fig. 2b show the measured plasmon frequency $\omega_p$ as a function of wave vector $q$ for monolayer $TaS_2$ (in cyan) and bilayer $TaS_2$ (in blue), demonstrating their flat dispersion relations. From the EELS spectra, we also measure an interband transition onset energy of 1.5 eV for both monolayer and bilayer samples (Fig. S7). This is the first experimental evidence that 2D plasmons in 2H-$TaS_2$ are well-separated from the electron-hole continuum, up to at least 0.15 $Å^{-1}$ without entering the electron-hole continuum where plasmons are strongly damped.

From the series of EELS measurements on the bilayer $TaS_2$, we can estimate the plasmon group velocity by taking the gradient of the dispersion relation as shown in Fig. 2a, resulting in a group velocity $v_g = d\omega/dq$ in the order of $10^{-4}c$ at the highest wave vector measured here $\sim 0.1 - 0.15$ Å$^{-1}$. This slow light behaviour leads to strong field confinement. The confinement factor can be estimated based on the maximum observable wave vector outside the electron-hole continuum. Here, we observe an in-plane confinement ratio of $\lambda_0/\lambda_p \sim 300$ for $q = 0.15$ Å$^{-1}$ at $\omega \approx 1\ eV$, exceeding the values <200 previously measured in graphene for Dirac plasmons[4] and edge plasmons[32,33].

Given the trend we observe towards stronger dispersion at high momentum, it is expected that the confinement ratio in $TaS_2$ will be larger—and consequently, the group velocity lower—for measurements done at even larger wave vectors. The maximum $q$ value measured here is simply limited by the sensitivity and stability of our experimental set-up, especially due to the low EELS cross-section in the monolayer limit compared to its bulk counter-part. At high $q$, the signal is also much lower than the unavoidable dark background in our used CCD-based EELS detector. Future measurements with single-electron sensitive EELS detectors[17] will be able to explore the whole Brillouin zone to find the onset of the e-h continuum where Landau damping dominates, and to confirm whether the negative dispersion relation that was observed in the bulk[11] persists in monolayers and few-layer $TaS_2$.

## Nonlocal interband screening

The observed slow light behaviour in $TaS_2$ arises from nonlocal (i.e. $q$-dependent) Coulomb screening, a universal feature of 2D materials[9,12,20,34–36]. Unlike bulk 3D materials where the dielectric screening is spatially uniform, the reduced dimensionality in 2D materials causes part of the electric field between charges to extend outside the material, making the screening strongly dependent on the distance ($r$)—or equivalently—the wave vector ($q$). For long-range interactions ($qd \ll 1$), screening is weak and dominated by the surrounding environment, whereas at short distances ($qd \gg 1$), it approaches the bulk limit. In 2D metallic $TaS_2$, this nonlocal screening is attributed to the polarizable background of the valence electrons.

Nonlocal interband screening can be characterized with an effective screening dielectric function $\varepsilon_{inter}(q)$ that screens the interaction with the free charge carriers in the plasmons[9,35,37,38]. Equation (2) describes the effect of interband screening $\varepsilon_{inter}(q)$ on the plasmon dispersion. In the low$-q$ limit ($qd \ll 1$), $\varepsilon_{inter}(q)$ can be represented analytically with the Keldysh model[39] for an ideal 2D monolayer[40]:

$$\varepsilon_{inter}(q) = 1 + \rho_0 q\ , \tag{4}$$

where $\rho_0 = d\varepsilon_{bulk}/2$ is the characteristic screening length. More accurate *ab initio* calculations (the striped line in Fig. 2c) show the deviation from the Keldysh model of $\varepsilon_{inter}(q)$ at large $q$ due to the finite thickness of the real 2D material[41].

Fig. 2c extracts the nonlocal screening in monolayers and bilayers of 2H-$TaS_2$ by comparing the measured screened dispersion relation with the $\sim\sqrt{q}$ dispersion relation of an ideal 2D electron gas. This procedure is described in detail in SI Section 5, with the extracted Drude weight and other fitting parameters listed in Table S1, and the robustness of the extracted results shown in Fig. S10. The extracted interband screening operates for both monolayer and bilayer $TaS_2$, and exhibits a clear momentum dependence, consistent with nonlocal dielectric screening in atomically thin metals. For the monolayer, the extracted screening length agrees well with *ab initio* calculations, validating the accuracy of the experimental model. As expected, the bilayer shows a higher degree of screening, attributed to its increased thickness, consistent with Equation (4), where the characteristic screening length $\rho_0$ scales proportionally with layer thickness. At large momentum values, both monolayer and bilayer datasets deviate from the linear dependence predicted by the idealized Keldysh model in Equation (4). This nonlinearity at high $q$ highlights a deviation from the ideal 2D electron gas, when the finite thickness of the layers becomes non-negligible and modifies the effective dielectric environment. The implication of finite thickness on plasmon behavior will be discussed in the following section.

## From 2D to 3D Coulomb interaction

In this section, we analyse the plasmon peak intensity as being the measured EELS probability density. As shown in Equation (3), this is defined by two terms: the loss function $\mathcal{L}(\omega, q)$—defined by the dielectric function of the material, as treated in the previous section—and the EELS prefactor $I_{kin}(q)$, which describes the interaction of the fast electron beam with the plasmons. In order to reconstruct the optical response of the sample material from measured EELS spectra through a Kramers-Kronig analysis, the EELS prefactor needs to be removed, as described in SI Section 2. Figs. 3a and 3b show the experimentally extracted loss function after removing the EELS prefactors $I_{kin}(q)\sim q^{-3}$ defined in Equation (S14). The simulated loss functions from the $TaS_2$ monolayer and bilayer are given in Figs. 3c and d, calculated from Equation (S12) with the fitted screening length and Drude weight extracted from Fig. 2b and Fig. S10.

The experimental loss function for the monolayer and bilayer 2H-$TaS_2$ is shown as orange background in Fig. 3a and Fig. 3b. It shows an unusual increasing trend at large wave vector $q$, different from the analytical model presented in Fig. 3c and Fig. 3d. The large mismatch between the experimental and theoretical loss function is due to the underestimation of the EELS prefactor at large $q$, which assumes a purely 2D Coulomb interaction in Equations (S11) and (S12). Clearly, an additional effect needs to be accounted for at large $q$ values. For a more complete description, we extract the EELS prefactor from the fitted Lorentz-Drude peak intensity in Equation (S16), with the experimentally extracted $\varepsilon_{inter}(q)$ shown in Fig. 2c. It now becomes clear that the EELS prefactors follow a $\sim q^{-3}$ relation in the low$-q$ regime and a $\sim q^{-2}$ relation for $q > 0.05$ Å$^{-1}$. Figs. 3e and 3f show these results, which are robust against different acquisition conditions (see Fig. S11), possible normalization artefacts (see Fig. S12) and reproducible across different samples (Fig. S13).

The extracted EELS prefactors for both monolayer and bilayer $TaS_2$ (Fig. 3e and 3f) reveal a clear transition from ideal 2D Coulomb interaction, characterized by a $\sim q^{-3}$ dependence, toward a $\sim q^{-2}$ scaling that is indicative of 3D-like behaviour at larger wave vectors. For the monolayer (Fig. 3e), this deviation becomes noticeable around $q \approx 0.03 - 0.05$ Å$^{-1}$, while the bilayer (Fig. 3f) shows a similar crossover at slightly lower values of $q \approx 0.02 - 0.04$ Å$^{-1}$. This trend is consistent with the expected dependence on material thickness: thicker films lead to an earlier breakdown of the 2D approximation. Notably, this onset coincides with the momentum range where the extracted interband screening $\varepsilon_{inter}(q)$ in Fig. 2c also begins to deviate from the linear relation predicted by the Keldysh model. This smooth but consistent departure from ideal 2D electron gas behaviour marks the onset of a dimensional crossover in the Coulomb interaction, where the finite thickness of the flakes begins to confine electric fields within the atomic layers[38]. Phenomenologically, this can be understood as a transition from a surface plasmon that decays with $\sim q^{-3}$[42,43] to a bulk plasmon that decays with $\sim q^{-2}$[42,44,45]. Essentially, for 2D electron gases in ultrathin films with vanishing thickness $d$ ($qd \ll 1$), most of the electric field lines are outside the material, resulting in a surface plasmon-like behaviour. At large $q$ however, the extent of the electric field in the out-of-plane direction is approximately $\lambda_p/2\pi \sim 0.6\ nm$, comparable with the layer thickness and the interlayer spacing. In this case, the field is confined almost completely within individual monolayers. As a result, the plasmons behave as bulk plasmons, consistent with the dispersionless property of bulk plasmons. This also means that plasmons in the high−$q$ regime are more sensitive to local defects and impurities in the 2D material.

It is important to distinguish the plasmons observed in atomically-thin $TaS_2$ from conventional surface plasmons in bulk metals. In bulk systems, surface plasmons are collective oscillations confined to the interface, with dispersion characteristics determined primarily by the dielectric mismatch across the interface. In contrast, the plasmons observed in atomically-thin $TaS_2$ satisfy $Re(\varepsilon) = 0$ and exhibit a Lorentz-Drude peak profile, as derived in SI Section 2—features that are conventionally associated with bulk plasmons. Therefore, plasmons in ultrathin metals such as $TaS_2$ combine characteristics of both surface and bulk plasmons: they are spectrally governed by bulk-like criteria but spatially extended and accessible similar to surface plasmons. These plasmons exhibit unique dispersion features such as a flat band and ultra-slow group velocity not typically observed in bulk counterparts. As the momentum increases, the field distribution transitions from being external (surface-like) to internal (bulk-like), marking a unique 2D-to-3D crossover in Coulomb interaction.

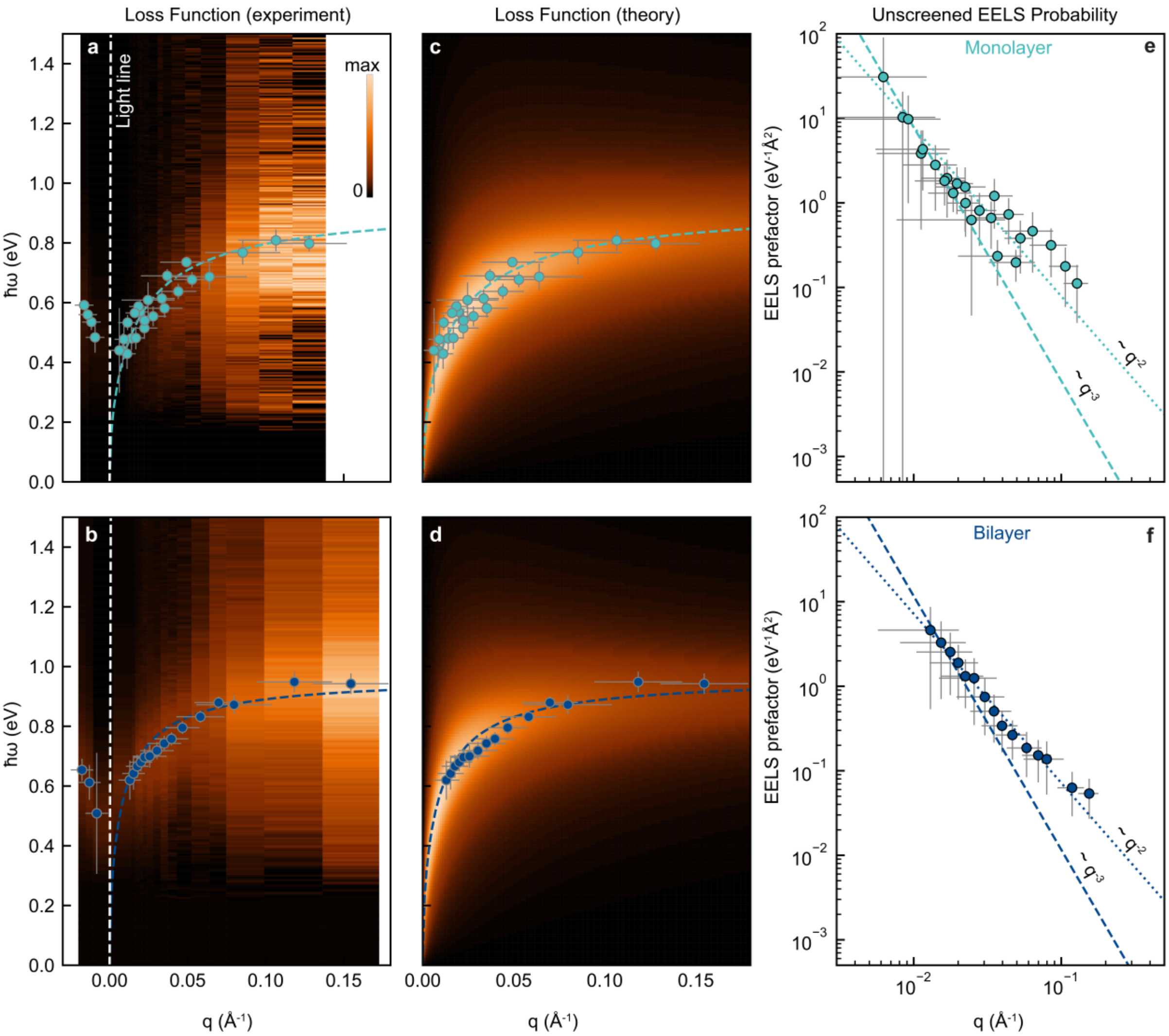

**Fig. 3 | Observing the transition from 2D to 3D Coulomb interaction. a** Experimental loss function (orange background) of 2H-$TaS_2$ monolayers and (**b**) bilayers, extracted from the measured plasmon peak intensity after removing the EELS prefactor $\sim q^{-3}$. **c** Simulated loss function of 2H-$TaS_2$ monolayer and (**d**) bilayer using Equation (S16) with fitting parameters from Fig. 2b and an estimated damping rate $\Gamma = 0.4$ eV. **e** Extracted EELS prefactor defined in Equation (S14) for 2H-$TaS_2$ monolayer and (**f**) bilayer from the fitted Lorentz-Drude amplitude after removing the interband screening factor $1/\varepsilon_{inter}(\boldsymbol{q})$ in Equation (3).

## Highly confined plasmons in suspended metallic monolayers

To further evaluate how plasmons propagate in $TaS_2$, we now quantitatively examine their damping rate from the width of the plasmon peaks. After removing the instrumental broadening from each measured spectrum[46] the inherent plasmon peak width $\Gamma(q)$ is obtained, presented in Fig. S14. As explained in SI Section 3b, the peaks broaden with decreasing wave vector due to the finite $\Delta q$ resolution. Besides this broadening at low $q$ predicted in Fig. S5, no trend in the peak width is observed in our measurements. $\Gamma(q)$ remains 0.3 – 0.4 eV, irrespective of the wave vector, indicating that there is no observable change in the damping mechanism and that the system does not enter the electron-hole-continuum regime up to the highest measured wave vector of 0.15 $Å^{-1}$. This conclusion is supported by the measured onset of the interband transition at 1.5 eV, shown in Fig. S7. These results underpin the initial hypothesis that the material can support plasmons up to a large wave vector without entering the lossy electron-hole continuum. From our measurements, we therefore conclude that highly-confined plasmons can be established in atomically-thin 2H-$TaS_2$ films.

Although the electron-hole continuum is avoided, the 2D plasmons in $TaS_2$ are still very lossy, with quality factors only up to about 3, in stark contrast with the sharp, highly underdamped plasmons calculated by da Jornada et al.[9] This suggests that mechanisms beyond intrinsic Landau damping are responsible for the observed losses. Here, we propose that *extrinsic* Landau damping, driven by atomic defects and especially phonons, plays a crucial role in 2D metals such as $TaS_2$. Landau damping is the energy transfer from the wave (plasmon) to the individual particles (electrons). Sometimes, this energy transition is forbidden by a mismatch in momentum when the plasmon is outside the electron-hole continuum. This momentum mismatch can be compensated via phonons or atomic defects, allowing energy dissipation through these extrinsic channels.

Extrinsic Landau damping was earlier demonstrated to be temperature-dependent in graphene[15,47]. Here we argue that this is also the case for $TaS_2$ monolayers and bilayers. First-principles calculations for monolayers[48] and optical measurements on bulk $TaS_2$[49] are used here to calculate the temperature-dependence of the peak widths, shown in Fig. S14. The lowest peak widths we experimentally observe are in fact comparable with the calculated electron-phonon relaxation time at room temperature, indeed indicating that phonon-assisted Landau damping is an important damping mechanism. In this regard, the calculated damping rates by Jornada et al.[9] are only relevant in the low temperature limit. At room temperature, thermal phonons remain the strongest source of inelastic plasmon scattering, while at lower temperature, thermal phonons are suppressed, making secondary damping pathways more pronounced, such as radiative losses, plasmon scattering on surface inhomogeneities due to oxidation and carbon residues (Fig. S15), and scattering on impurities and crystal defects[50]. We also considered other damping pathways and excluded them. For example, the excitation of electron-hole pairs is inefficient[51], and direct plasmon-phonon coupling is weak in this case due to the large mismatch between the plasmon energy around 1 eV and the phonon energy around 0.05 eV, shown in the Raman spectra of Fig. 1c. The dose effect of the electron beam is also excluded, as it does not result in significant changes in the plasmon peak, as shown Fig. S16.

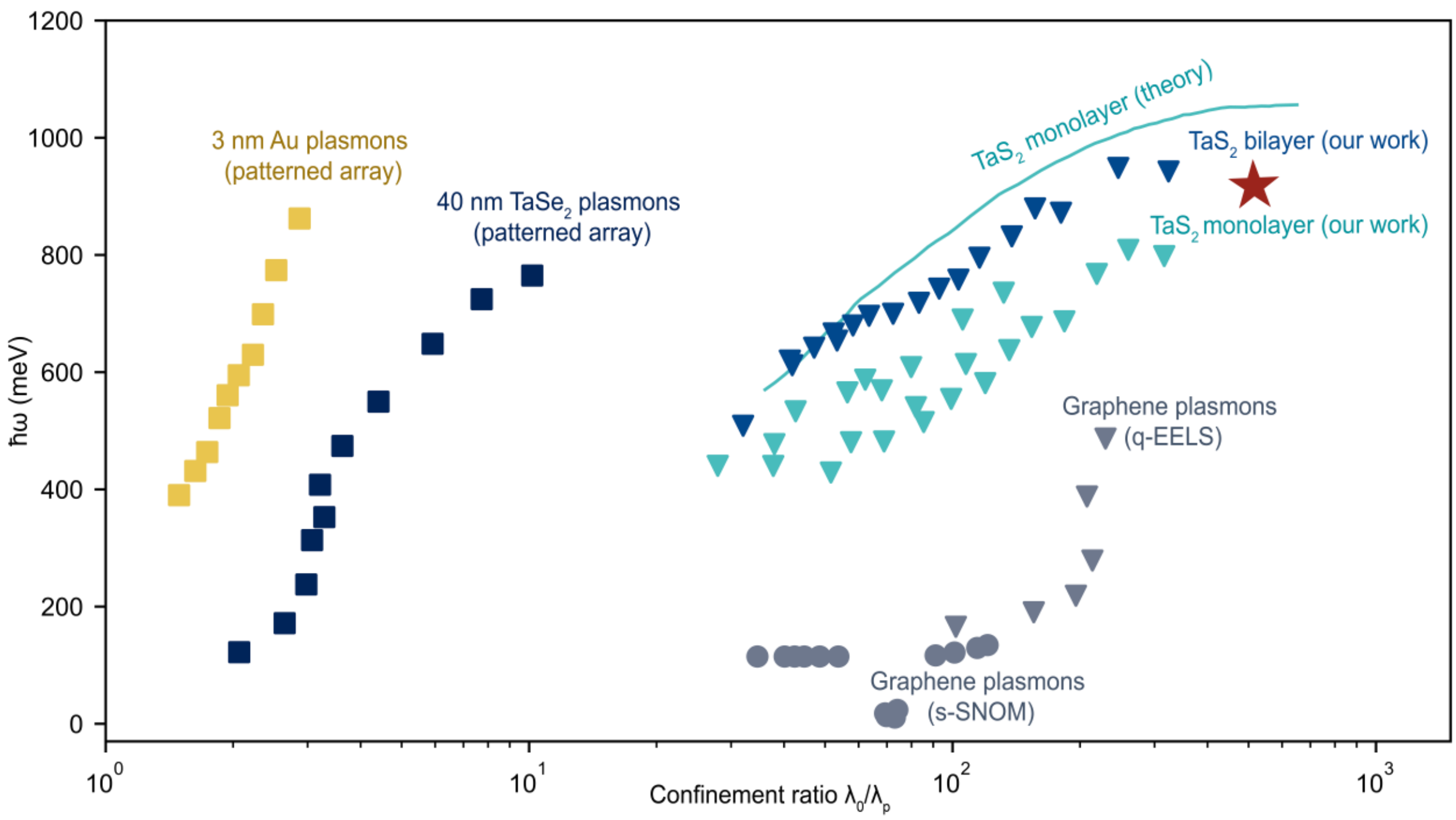

**Fig. 4 | Plasmon confinement for various thin films.** The dispersion relation of 2D plasmons and polaritons of various thin films are shown, as a function of their confinement ratio. Patterned arrays of 3 nm thick gold and 40 nm-thick $TaSe_2$ were measured with FTIR[12,52]; Dirac plasmons and acoustic plasmons in graphene were measured with s-SNOM[53,54] and Dirac plasmons in graphene on polar substrates were measured by reflective q-EELS[6]. The results presented in this current work are indicated by the red star, showing the highest plasmon confinement ratios measured so far.

# Conclusions

In this work, we experimentally confirmed the theoretical predictions that plasmons in monolayer and bilayer 2H-$TaS_2$ disperse to large wave vector values outside the electron-hole continuum. This marks the first direct experimental evidence of nonlocal screening in 2D metals, demonstrating that the origin of dispersionless plasmons lies in the light field being tightly confined within the atomically-thin films. Additionally, our findings confirm the predicted slow light behaviour and extreme light confinement down to the scale of 1-2 nm. Our results compare positively with reported confinement ratios in other ultrathin metallic films, as summarized in Fig. 4.

This work therefore demonstrates that $TaS_2$ and 2D metallic TMDs in general can be potential platforms for flat-band photonics[55] and deep sub-wavelength optical spectroscopy applications in the near-infrared regime. The confinement ratio surpasses that of 2D plasmons in graphene and is even on par with phonon-polaritons in the mid-infrared regime[17].

Future work at cryogenic temperatures will be able to quantify the reduction in phonon-assisted Landau damping that we predict, as well as the effect of charge density waves on plasmon scattering. Investigations on encapsulated monolayers and multilayers will provide further insight into plasmon coupling and their scattering on structural defects. The experimental and analytical routines we presented here for quantitative q-EELS are applicable to any other (semi-)metallic 2D materials, offering a flexible approach for nano-optical spectroscopy without substrate effects and with high q-resolution.

# Materials and Methods

## Electrochemical exfoliation

A top-down method was used to exfoliate monolayer 2H-$TaS_2$ from a bulk crystal (purchased from HQ graphene). It is particularly more difficult to obtain monolayers of Ta-based TMDs using mechanical cleavage[56,57] than from semiconducting TMDs. In this work, we therefore adopt electrochemical exfoliation, to obtain large mono- and few-layer $TaS_2$ flakes with high yield. The liquid also protects the monolayer from oxidation during the exfoliation process, as monolayer and few-layer $TaS_2$ flakes are sensitive to oxidation[58]. Our electrochemical exfoliation procedure follows the procedure previously described in Zhao et al.[25]: tetrapropyl ammonium tetrafluoroborate in propylene carbonate was used as the electrolyte , with the bulk $TaS_2$ crystal as the cathode and Pt wire as anode. After exfoliation, the solution was centrifuged at 2500 rpm to remove the unexfoliated and thick flakes. The supernatant left with ultrathin flakes was collected and further washed at 13000 rpm for three times. If there were still thick flakes, centrifugation at a higher speed would be carried out to get high yield of monolayer and bilayers. The final sediment can be redispersed in various solvents, such as PC, DMF and IPA, with small volume, so that a final dispersion of $TaS_2$ monolayers with high yield and high concentration was obtained. The final dispersion would go through a mild sonication to break the large flakes into smaller pieces, for the ease of dry transfer during TEM sample preparation; the latter process is described in SI Section 1.

## Crystal structure and electrical transport measurements

Device fabrication for superconductivity and electrical measurements was performed using standard electron beam lithography and e-beam deposition of Ti/Au (5/50 nm) electrodes. The devices were encapsulated with h-BN flakes in an inert glovebox. We carried out four-terminal resistance measurements with an SR830 lock-in amplifier in an Oxford TeslatronPT system operating at a base temperature of 1.5 K.

XRD measurements were performed at room temperature using a Bruker Kappa-APEX2 single crystal diffractometer, equipped with sensitive Apex II CCD detector using a copper X-ray source. The XRD results are shown in Fig. S1. Only four peaks are observed in the entire spectrum with sharp lines, indicating the high crystallinity. It agrees well with the pure 2H phase XRD data (PDF 01-080-0685), with the four peaks indexed as (00l) reflections, with l = 2, 4, 6, and 8.

Raman spectroscopy confirms the phase and sample quality of the samples. The spectra were collected at room temperature using a WITec Alpha300 R photon scanning tunnelling microscope (PSTM) equipped with a 532 nm laser. Single spot spectra were collected under a 100× objective and the laser intensity was optimised to get a clear signal from monolayer samples without damaging them. After optimisation, the laser intensity was kept the same for all the measurements.

Monolayers and bilayers were confirmed from optical contrast with reflective bright field light microscopy, and double-confirmed in the STEM through the bulk plasmon peak as shown in Fig. S3. The exfoliated flakes are transferred onto holey SiN TEM grids as described in SI Section 1.

## Momentum-resolved EELS

STEM-EELS measurements were done using a non-aberration-corrected ThermoFisher Titan TEM with a Schottky electron source operated at 80 kV. A single Wien-type monochromator dispersed the electron beam in energy, and a narrow energy-selecting slit formed a monochrome electron beam with typical full-width at half-maximum values of ~50 meV. Our Titan TEM is a first-generation monochromated system, with a post-column Gatan Tridiem EELS detector. This system does not provide the convenience of performing q-EELS using a modern ω-q slit that would allow fast, parallel acquisition. Instead, our q-EELS experiments were performed in a serial manner using free-lens control to perform precise shifts in momentum-space for each spectrum measurement. Following a method described earlier[21,22], we used a convergence semi-angle of 1.38 mrad and physically pushed the sample furthest out of focus (~300 µm) to obtain a continuously magnified diffraction pattern on the Tridiem camera. The irradiated area of the sample was about 500 nm in diameter. The projector lenses magnified the diffraction pattern by 10000-27000 times, resulting in an estimated camera length of about 4-8 m (Fig. S4). We varied the projector magnification to control the EELS collection angles, as shown by the different sizes of the green disks in Fig. 2a. For each measurement, the projector shift was used to allow different positions in k-space to enter the 1 mm entrance aperture of the EELS detector. Despite the lack of a ω-q slit, it is possible to obtain a high momentum resolution, up to q=0.0056 $\text{Å}^{-1}$, limited by the $z$-height range, sample charging, optical stability and the size of the smallest EELS entrance aperture. The total momentum resolution is presented as the error bars for wave-vector $q$ in all experimental data points presented above. For each q value, a series of 2000 individual, raw (unprocessed) spectra were collected, each with an exposure time ranging from 50 ms to 200 ms using the StripeSTEM routine[59], and applying binned gain averaging[60]. Separately, 2000 dark background spectra were collected with the same StripeSTEM routine, but now with the electron beam blanked. The average dark background spectrum was subtracted from all raw lowloss spectra, subsequently aligned and summed to finally obtain one EELS spectrum with very high signal-to-noise values.

The momentum resolution $\Delta q$ of our error bars also includes the Heisenberg uncertainty limit $\Delta q_{\beta}$ defined by the collection angle, and the Abbe diffraction limit $\Delta q_{\alpha}$ defined by the convergence angle. This setup provides us with a flexibility to continuously vary the camera length, and therefore $\Delta q_{\beta}$ through defocus $z$ and the magnification from the projector lenses, as shown in SI Section 3.

Aberration corrected STEM images (Fig. S2) were acquired with a JEOL ARM200CF equipped with a cold field emitter and DCOR aberration corrector[61] [Loh 2025], operating at 60 kV. An electron beam convergence semi-angle of 31 mrad was used, and electrons that were forward-scattered between 68 and 280 mrad were measured for atomic-resolution annular dark field imaging.

# Acknowledgements

H.T.B.D. and MB kindly acknowledge support from the Singapore Ministry of Education via the Academic Research Fund (project number MOE-T2EP50122-0016). P.S. thanks the Singapore Ministry of Education for support under grant RG113/21. M.Z. kindly acknowledges the Agency for Science, Technology and Research (A*STAR) under its Career Development Fund C210812027. J.H.T. acknowledges funding support from the National Research Foundation Singapore under the CRP program (Grant No. NRF-CRP26-2021-0004) and A*STAR for HBMS IAF-PP Grant No. H19H6a0025. Markus Heidelmann (University of Duisburg-Essen) is kindly acknowledged for providing the StripeSTEM routine. We are grateful for the technical support from Ms. Teo Siew Lang and Mr. Lim Poh Chong from A*STAR.

# Author contributions

H.T.B.D., M.Z., and M.B. conceived the research. M.Z. together with P.L., Y.W.S., J.L., and J.H.T. performed electrochemical exfoliation, post treatment, and optical characterization. J.R., H.T.B.D.,

M.Z., and X.Z. performed TEM sample preparation. B.L., T.J., and P.S. fabricated the electrical devices and performed superconductivity measurements. H.T.B.D. designed & conducted the EELS experiments under the supervision of M.B. H.T.B.D. performed the EELS data analysis, analytical modelling, and prepared the original draft, which was edited and reviewed by M.Z. and M.B., incorporating comments from all authors.

# Competing interests

The authors declare no competing interests.

# Materials & Correspondence

Correspondence to Meng Zhao (zhaom@imre.a-star.edu.sg) and Michel Bosman (msemb@nus.edu.sg).

Supplementary Information for

# Slow, Nanometer Light Confinement Observed in Atomically Thin $TaS_2$

Hue T.B. Do, Meng Zhao*, Pengfei Li, Yu Wei Soh, Jagadesh Rangaraj, Bingyan Liu, Tianyu Jiang, Xinyue Zhang, Jiong Lu, Peng Song, Jinghua Teng, Michel Bosman*

*zhaom@imre.a-star.edu.sg; *msemb@nus.edu.sg

# 1. Sample preparation and thickness confirmation

## a. Monocrystalline 2H-$TaS_2$ measured by X-ray diffraction

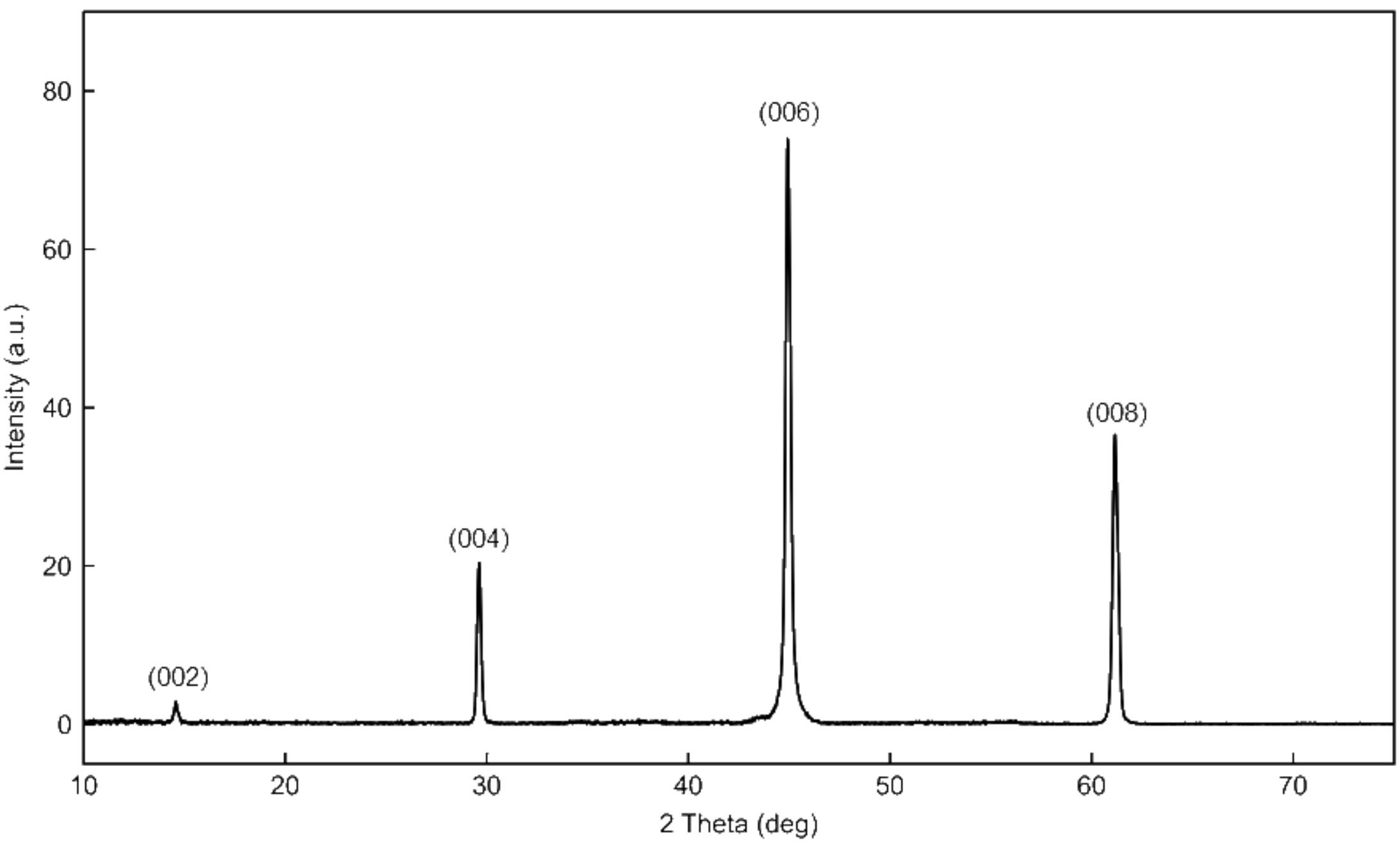

*Fig. S1 | Single-crystal XRD from the $TaS_2$ crystal that was subsequently exfoliated for the creation of monolayers and bilayers. The four peaks agree well with the pure 2H phase XRD data (PDF 01-080-0685), indexed as (002), (004), (006), and (008) reflections.*

## b. TEM sample preparation and thickness confirmation

The suspension of $TaS_2$ flakes in propylene carbonate is drop-cast and vacuum dried on a $SiO_2$/Si substrate where the monolayers and bilayers are identified based on their optical contrast[1,2]. The selected flakes are then transferred onto an Au-coated, amorphous $SiN_x$ holey TEM grid through the polycarbonate transfer technique as described earlier[3] and washed with chloroform in an inert environment. This leaves free-hanging monolayer and few-layer $TaS_2$ flakes supported on Au/SiN. The described Au coating is needed to avoid charging effects of the electron beam during EELS measurements. Then the TEM grid is loaded in a TEM sample holder. The entire procedure is done in the glovebox to avoid oxidation, except for a few seconds in air while loading the holder in the TEM. Fig. S2 shows representative atomic-resolution images of the transferred $TaS_2$.

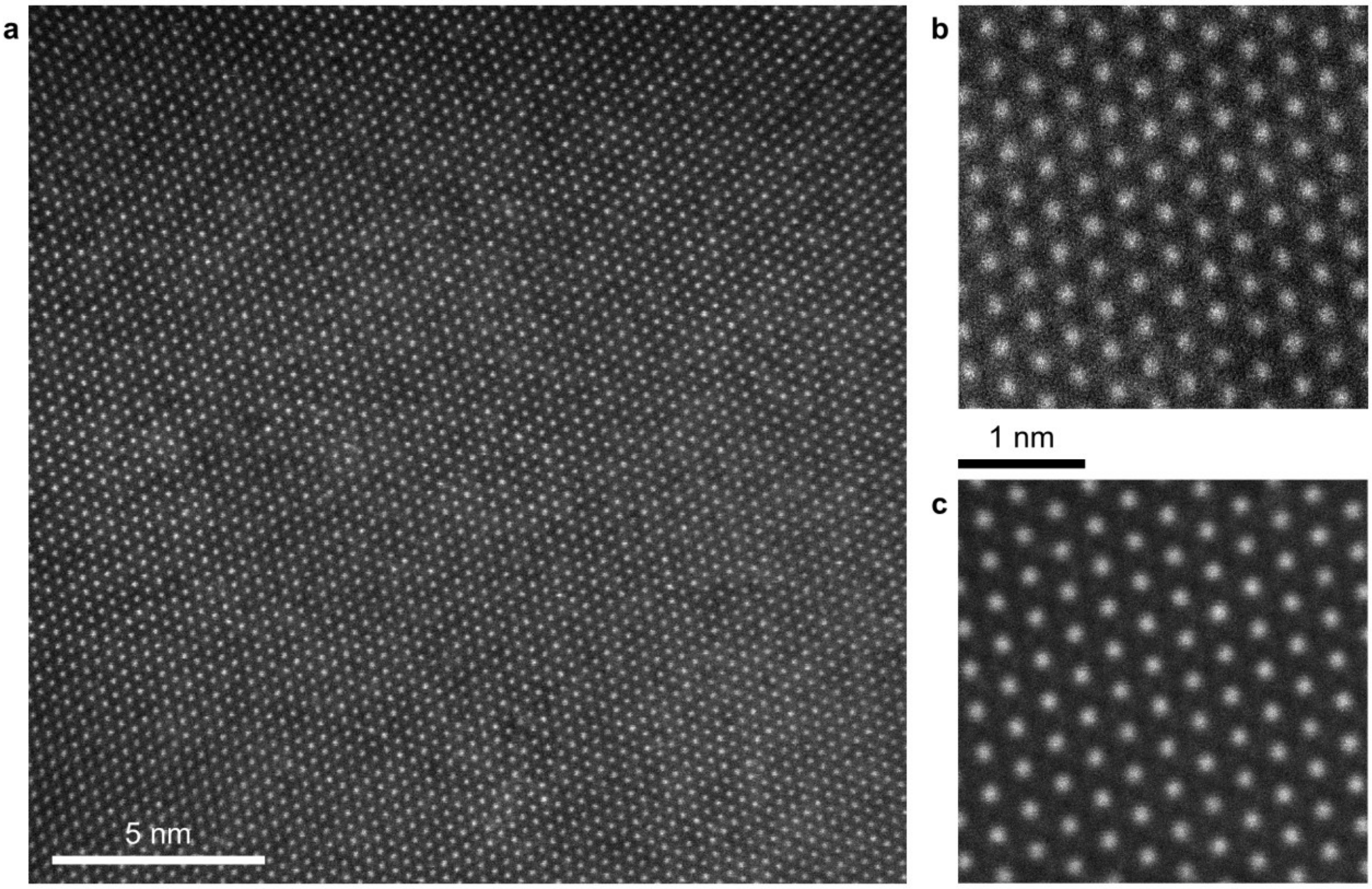

*Fig. S2 | Aberration-corrected, high-angle annular dark field STEM images of the freely-suspended 2H-$TaS_2$ films that were used for the momentum-resolved EELS experiments. a) Large-area view of the $TaS_2$ bilayer. b) Detail of the $TaS_2$ monolayer and of the c) $TaS_2$ bilayer. Images were acquired at 60 keV, using a JEOL ARM200CF with a UHR polepiece and CEOS ASCOR aberration-corrector.*

Differentiating 2H-$TaS_2$ monolayer from bilayers on TEM grids is done *via* the optical contrast of the flake on the Au/SiN TEM grids (Fig. Sa, b) and further confirmed through the π +σ plasmon peak position around 14-16 eV in the low-loss EELS spectrum. Fig. Sc shows a typical low-loss EELS spectrum of two sampled 2H-$TaS_2$ flakes. We observe a $TaS_2$ bulk plasmon peak of all electrons (π and σ) at 14.6 ± 0.4 eV and 16.8 ± 0.5 eV, much lower than previously reported in bulk sample (at 21 eV)[4]. The lower bulk plasmon energy of a 2D material compared to its bulk counterpart is expected, and has been shown in EELS measurement of other 2D materials such as graphene[5], boron nitride[6] and $MoS_2$[7]. Therefore, we can use the bulk plasmon peak as a fingerprint for thickness determination. This is especially useful for oxidation-prone Ta-based TMDs, as the alternative method for thickness determination, AFM, is usually conducted in ambient conditions.

We compare the bulk plasmon peak energies to those obtained from DFT calculations[8] to determine whether the flake is a monolayer or bilayer (Fig. Sd). The bulk plasmon (π+σ) here is attributed to the π–σ* and σ–σ* transition, while the collective π–π* transition (π plasmon) at <9 eV is difficult to distinguish from that of propylene carbonate, the solvent used for exfoliation. The peak at 3 eV is attributed to the onset of an interband transition and consistent with previous reports in bulk samples.

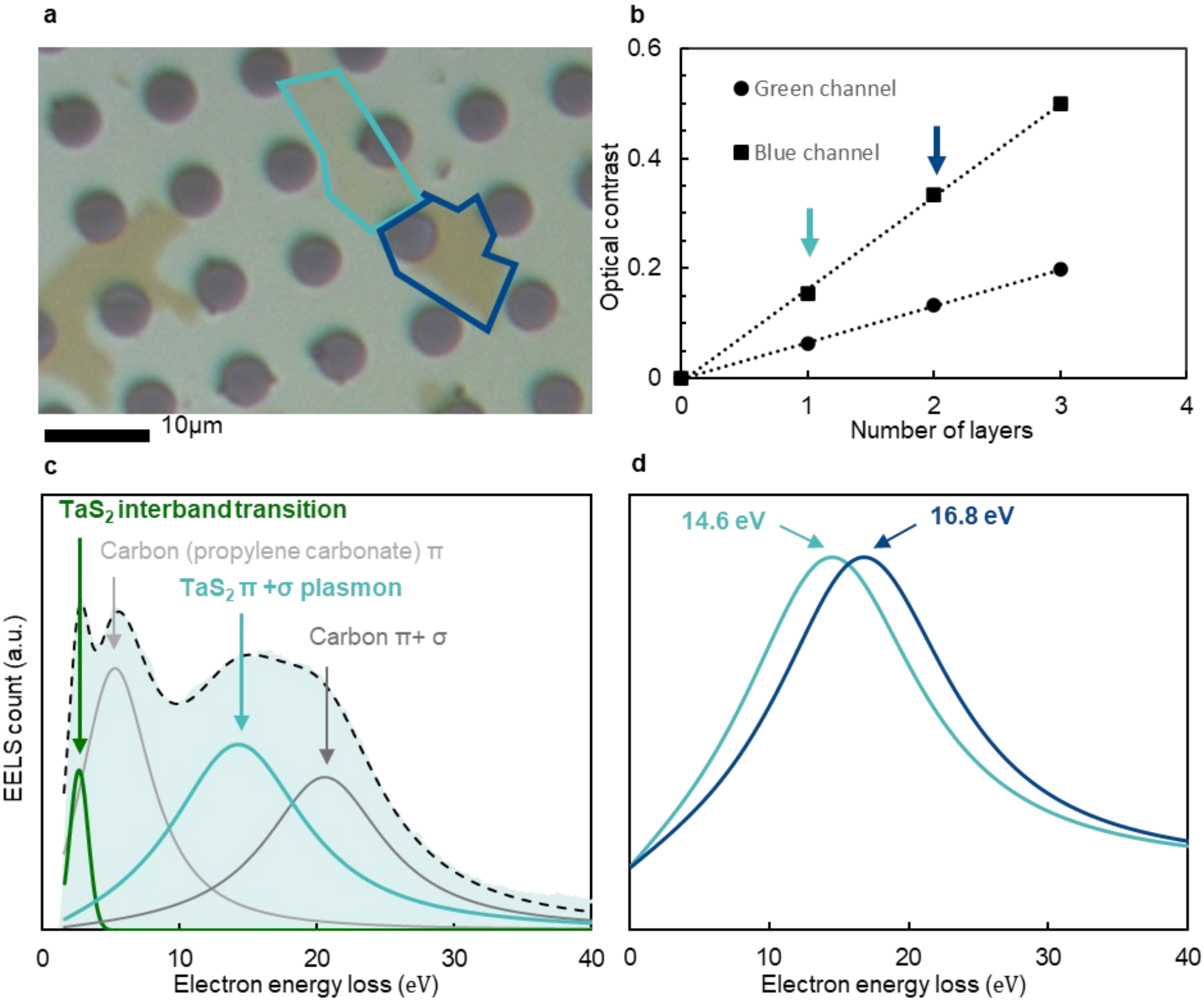

*Fig. S3 | a) The 2H-$TaS_2$ TEM sample is prepared though a polymer-assisted transfer onto an Au-coated holey SiN TEM membrane. b) Thickness-dependent optical contrast of the 2H-$TaS_2$ on a perforated Au/SiN membrane. c) Low-loss EELS measurement on monolayer sample with overlaid the fitting for the $TaS_2$ interband transition (more details in Fig. S6), the π and π+σ signal from propylene carbonate, and the π+σ $TaS_2$ plasmon. d) π+σ plasmon peaks of monolayer (cyan) and bilayer (blue) $TaS_2$, fitted from measured low-loss EELS spectra. The low-loss spectra are acquired at ultra-small collection angle in the momentum-resolved EELS setup (integrated from q=0 to $q=q_{max}$ with $q_{max}=0.18$ Å$^{-1}$ for the monolayer data and $q_{max}=0.05$ Å$^{-1}$ for the bilayer data). The peak positions agree with previous DFT calculations[8].*

## 2. Theoretical formalism of q-EELS

### a. 2D dielectric function

In this section, we derive the dielectric function of 2H-$TaS_2$ based on the random-phase-approximation framework adopted from graphene. As suggested by da Jornada et al.[9], the optical response of a $TaS_2$ monolayer can be represented as an ideal 2D electron gas embedded in a dielectric environment accounting for interband screening and substrate effects. A similar approach was implemented to model 2H-$TaSe_2$ thin films (>10 nm) by Song et al.[10] in the long wavelength limit. The electron energy loss probability of the electron beam to excite a 2D plasmon polariton is calculated within the quasi-static approximation for an infinitely thin 2D system, adapted from Rodríguez Echarri et al.[11] for atomically-thin noble metal films.

We first consider the case of an ideal 2D electron gas (2DEG) where electrons are confined to move only in the in-plane (longitudinal) direction with a sheet of electron density $n$. The longitudinal sheet conductivity of this 2D electron gas in the Drude model within the linear and local response is:

$$\sigma(\omega) = \frac{e^2 n}{m} \frac{1}{\Gamma - i\omega} \; . \tag{S1}$$

Adopting from the random-phase-approximation (RPA), the dielectric function can be written in terms of the Lindhard polarization[12].

$$\varepsilon_{2D}(q,\omega) = \varepsilon_r - \nu_q P(q,\omega), \tag{S2}$$

where $\varepsilon_r$ is the dielectric function of the embedded medium, $\nu_q = \frac{e^2}{2\varepsilon_0 q}$ is the spatial Fourier transform of the Coulomb interaction in 2D and $P(\boldsymbol{q},\omega)$ is the Lindhard polarization. The relation between the 2D conductivity and polarization has been shown[12] to be the following:

$$\sigma(q,\omega) = ie^2 \frac{\omega}{q^2} P(q,\omega). \tag{S3}$$

We can then write the dielectric function of this 2DEG as:

$$\varepsilon_{2D}(q,\omega) = \varepsilon_r - \frac{e^2 n}{2\varepsilon_0 m} \times \frac{1}{\omega(\omega + i\Gamma)} q. \tag{S4}$$

The term $\mathcal{D} = \frac{\pi e^2 n}{m}$ is defined as the Drude weight of the 2DEG. Damping pathways are phenomenologically expressed through the scattering rate $\Gamma$. Damping pathways here include intrinsic interband and intraband Landau damping as well as extrinsic scattering on phonons and impurities. Note that the inclusion of the damping term $\omega + i\Gamma$ here agrees with the relaxation-time approximation by Mermin[13] that has been used for the case of graphene[14].

For real 2D metals, it is conventional to include a series of Lorentz oscillator terms representing each interband transition. For $TaS_2$, da Jornada et al.[9] have shown that, for small $q$ up to $0.4\ \text{Å}^{-1}$, the contribution of interband screening can be expressed as $\varepsilon_{inter}(q,\omega) = 1 + \rho_0 q$ , where $\rho_0$ is the 2D screening length, which can be obtained from first-principle calculations. The material can now be represented as an ideal 2DEG embedded in an effective screening environment $\varepsilon_{inter}(q,\omega)$. In our case, there is no substrate effect as 2D flakes are free-hanging, suspended in the vacuum of the STEM. The effective dielectric function becomes:

$$\varepsilon_{2D}(q,\omega) = \varepsilon_{inter} - \frac{\mathcal{D}}{2\pi\varepsilon_0} \times \frac{1}{\omega(\omega + i\Gamma)} q \,. \qquad (S5)$$

Rewriting Equation (S5) in terms of

$$\omega_p = \sqrt{\frac{\mathcal{D}}{2\pi\varepsilon_0} \times \frac{q}{\varepsilon_{inter}}}, \qquad (S6)$$

we obtain the dielectric function in the familiar Lorentz-Drude form:

$$\varepsilon_{2D}(q,\omega) = \varepsilon_{inter}\left(1 - \frac{{\omega_p}^2}{\omega(\omega + i\Gamma)}\right). \qquad (S7)$$

Equation (S6) becomes the dispersion relation of 2D plasmons in $TaS_2$ as shown in previous calculations. This corresponds to the zeros of the $Re(\varepsilon_{2D}(q,\omega))$ under negligible damping approximation[15], or through applying the dispersion relation for an ideal 2D electron gas with conductivity $\sigma(q,\omega)$ from Equation (S1) embedded in a dielectric environment $\varepsilon_{inter}$.

### b. EELS probability

In an ω-q EELS experiment, the value of transferred momentum $q$ can be obtained from a selected scattering angle $\theta$ of the incoming electron beam after interaction with the 2D metals. Here we present the relation between in-plane transferred momentum $q$ and scattering angle $\theta$ through conservation of momentum and energy[16]. Relativistic treatment of the incident electron beam is considered as we use 80 kV incident electrons, equivalent to $v \approx 0.5c$, with a relativistic factor $\gamma = \frac{m^*}{m_0} = \frac{1}{\sqrt{1-v^2/c^2}} = 1.155$.

Considering an inelastically scattered electron with initial total energy $W_i = \gamma m_0 c^2$ and initial momentum $p = \gamma m_0 v = \hbar k_i$, its initial energy and momentum are related through

$$W_i = \sqrt{m_0 c^2 + (\hbar k_i c)^2} \,. \qquad (S8)$$

After going through an energy loss $E = \hbar\omega$ to the 2D metals, we have a similar relation between the final energy and momentum:

$$W_f = W_i - \hbar\omega = \sqrt{m_0 c^2 + \left(\hbar k_f c\right)^2} \,. \qquad (S9)$$

Combining Equation (S8) and (S9), we can write the final momentum $k_f$ in terms of $k_i$ and $\omega$ and independent of scattering angle $\theta$, with an approximation $E = \hbar\omega \ll W_i$ to omit the $E^2$ term:

$${k_f}^2 = {k_i}^2 - \frac{2\gamma m_0 \omega}{\hbar} \qquad (S10)$$

Conservation of momentum gives the two components of transferred momentum: $q_\perp = k_f sin\theta$ and $q_\parallel = k_i - k_f cos\theta$. The momentum of the surface plasmon polariton in 2D plasmons is purely in-plane and perpendicular to the electron path. Hence the momentum $q$ used in plasmon dispersion is $q = q_\perp = k_f sin\theta = k_i sin\theta\sqrt{1 - \frac{2\hbar\omega}{\gamma m_0 v^2}}$. In our spectral region of interest with $q \sim 0.1$ Å$^{-1}$ and $\hbar\omega \sim 1\ eV$, we can estimate that $\frac{2\hbar\omega}{\gamma m_0 v^2} \sim 10^{-5} \ll 1$ and $sin\theta \sim 10^{-3} \ll 1$. Therefore, we can safely assume $q_\perp \approx$ $k_i\theta$ and $q_{\parallel} \approx \frac{\omega}{v}$. We can then use the diffraction pattern of the elastically scattered electrons to calibrate the momentum space of inelastic electrons as shown in SI Section 3.

The EELS spectrum collected in each green disk in Fig. 2a represents the probability density $\frac{d^2P}{d\Omega dE}$ of the energy loss $E = \hbar\omega$ and the in-plane momentum transfer $\hbar\vec{q}$ in a solid angle $d\Omega = 2\pi q dq$. In this section, we derived the probability density $\frac{d^2P}{d\Omega dE}$ for the case of 2D electron gas with vanishing thickness $d$ ($qd \ll 1$) in the non-relativistic limit with negligible radiation loss ($\omega \ll cq$).

An analytical model for the scattering probability can be obtained in the nonrelativistic and quasistatic limit, and we adopt the formulation of Echarri et al.[11] for EELS probability of a-few-atomic-layer thin metallic film (Au and Ag), in the case of perpendicular electron trajectory:

$$\frac{d^2P(q,\omega)}{dqd\omega} = \frac{8e^2}{\hbar v^2} \times \frac{q}{\left(q^2 + \frac{\omega^2}{v^2}\right)^2} \int dz \int dz' cos\,[\frac{\omega(z-z')}{v}]$$

$$\times\, Im\{-\chi(q,z,z',\omega)\}\ (Gaussian\ unit) \qquad (S11)$$

with $\chi(q,z,z',\omega)$ as the linear susceptibility, and $q$ describing the in-plane momentum.

Assuming an ideal 2D electron gas in van der Waals 2D metals, with the electron wavefunction in the out-of-plane direction, ( $z$ ) is in the form of a Dirac Delta function, so we can write: $Im\{-\chi(q,z,z',\omega)\} = Im\{-\chi(q,\omega)\} \times \delta(z) \times \delta(z')$. $\chi(q,\omega)$ can be obtained from $\varepsilon^{-1}(q,\omega) = 1 + v_q\chi(q,\omega)$, so we obtain[17]:

$$\frac{d^2P}{d\Omega dE}\ (\omega,q) = \frac{e^2}{2\pi^2\varepsilon_0\hbar^2v^2} \times \frac{q}{\left(q^2 + \frac{\omega^2}{v^2}\right)^2} \times Im(\varepsilon_{2D}^{-1}(\omega,\boldsymbol{q})) \qquad (S12)$$

The derived EELS probability agrees with previous formalisms for the case of graphene[18–22] excluding radiation loss. Radiation loss is indeed negligible, especially in the flat dispersion regime as the plasmon disperse with group velocity $v_g \ll c$ and far away from the light line in the measured regime. This formalism is valid for the long wavelength limit ($qd \ll 1$) and diverges at large $q$ as shown in the discussion of Fig. 3.

Equation (S12) contains 2 parts:

a) An EELS prefactor, describing the Coulomb interaction between the fast-moving electron and the sample material:

$$I_{kin}(\omega,q) = \frac{e^2}{2\pi^2\varepsilon_0\hbar^2v^2} \times \frac{q}{\left(q^2 + \frac{\omega^2}{v^2}\right)^2} \qquad (S13)$$

The formalism results in a similar expression in the limit of vanishing thickness as previously shown in thin films[23,24].

For $v = 0.5c$ being the speed of the incoming electron beam, $q \gg q_{\parallel} = \omega/v$, the prefactor can be simplified as

$$I_{kin}(q) = \frac{e^2}{2\pi^2\varepsilon_0\hbar^2v^2} \times \frac{1}{q^3} \qquad (S14)$$

In passing, $q_{\parallel}$ being the out-of-plane momentum transfer has been shown to result in quenching of the EELS intensity[25] in the very long wavelength limit $q \to 0$.

b) The loss function of the sample material, depending on its dielectric function:

$$\mathcal{L}(\omega, q) = Im(\varepsilon_{2D}^{-1}(\omega, q)) \quad (S15)$$

Using the 2D effective dielectric function in Equation (S7) the loss function for free-hanging 2H-$TaS_2$ monolayers and bilayers can be written as a Lorentz-Drude function:

$$\mathcal{L}(\omega, q) = \frac{1}{\varepsilon_{inter}(q)} \times \frac{\omega \Gamma \omega_p(q)^2}{\left(\omega^2 - \omega_p(q)^2\right)^2 + (\omega \Gamma)^2} \quad (S16)$$

For each EELS spectrum, we remove the quasi-elastic background (Fig. S6) and fit the plasmon peak with a Lorentz-Drude function. From this, we obtain for each spectrum a specific set of fitting parameters:

(1) Peak position $\omega_p(q) = \sqrt{\frac{\mathcal{D}}{2\pi\varepsilon_0} \times \frac{q}{\varepsilon_{inter}(q)}}$,
(2) Peak intensity $A(q) = I_{kin}(q) \times \frac{1}{\varepsilon_{inter}(q)} = \frac{e^2}{2\pi^2 \varepsilon_0 \hbar^2 v^2} \times \frac{1}{q^3} \times \frac{1}{\varepsilon_{inter}(q)}$, and
(3) Peak width $\Gamma(q)$.

## 3. Momentum-resolved EELS setup and calibration

### a. Momentum calibration

As described in the Methods section, the sample is pushed out-of-focus with defocus $z = 338\ \mu m$. At this defocus, the diffraction pattern is much larger than the camera collection angle. We extrapolate the momentum calibration using a simple geometric relation shown in Midgley[26]. The effective camera length is proportional to the defocus as shown in Fig. S4.

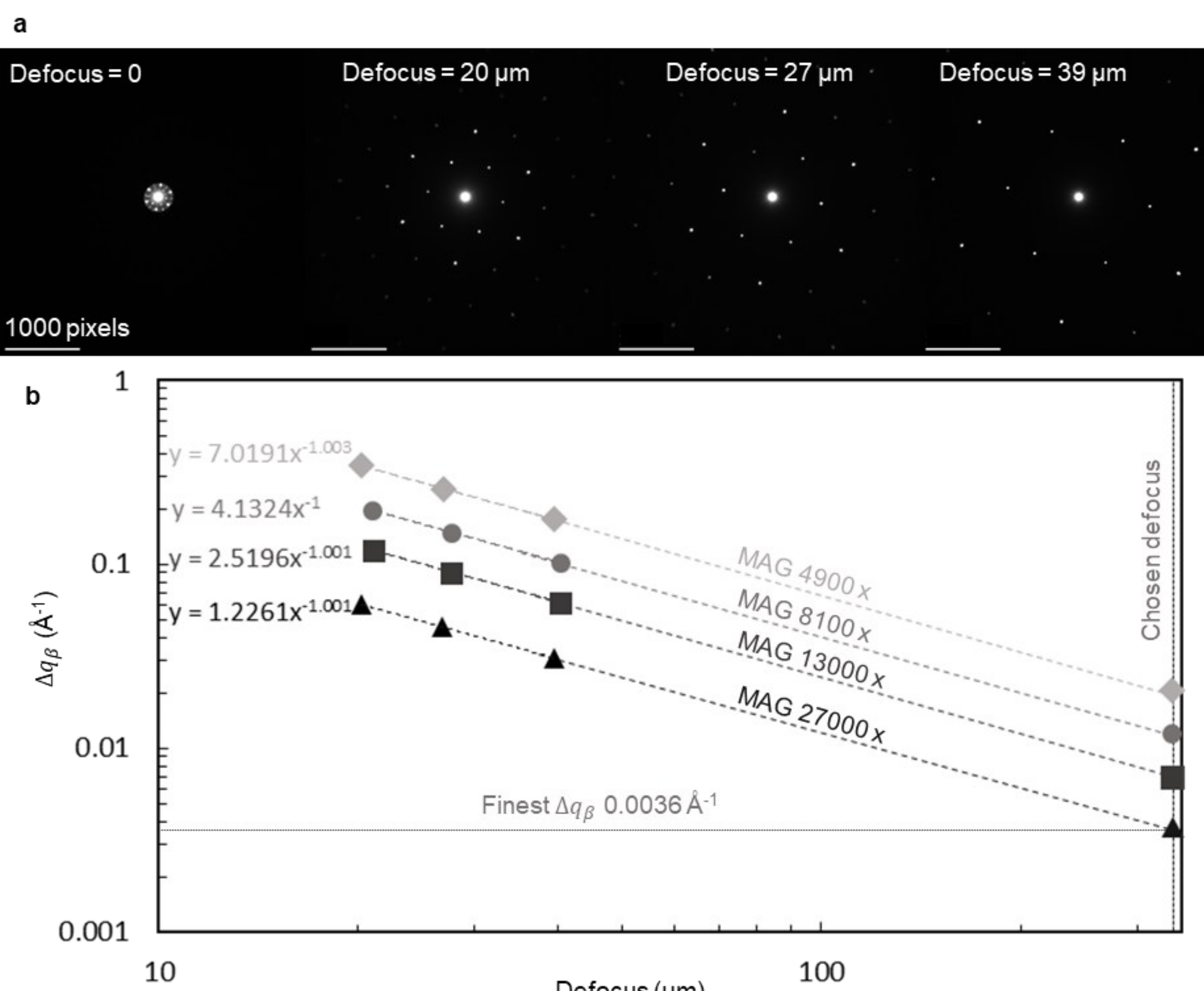

*Fig. S4 | The q-resolution can be tuned by both defocus and magnification of the projector system. This provides flexibility in optimizing the q-resolution (momentum space) and the illuminated area (real space). a) The camera length can be tuned continuously by changing the defocus[26,27]. b) Consolidated q-resolution as a function of defocus and magnification of the projection system for our TEM. This calibration is done for a convergence angle of 1.38 mrad.*

### b. Momentum-resolution evaluation

As discussed in the main text, the effect of finite momentum resolution is significant and needs to be treated cautiously in interpreting the EELS results. The momentum resolution $\Delta q$ in this work is limited by the physical probe size $d_\alpha$ and the size of the EELS entrance aperture $d_\beta$. With a small convergence angle ($\alpha \sim 1.38\ mrad$) used in this work, the lens aberrations are negligible, and $d_\alpha$ is diffraction limited:

$$d_\alpha = \frac{\lambda}{2 \sin \alpha} \qquad (S17)$$

The contribution of the probe size to the momentum resolution is

$$\Delta q_\alpha = \frac{d_\alpha}{z} k_i = \frac{\pi}{\mathrm{z} \sin \alpha} \qquad (S18)$$

with $k_i$ being the momentum of the incoming electron beam. Typical $\Delta q_\alpha$ in our experiments is $0.002$ Å$^{-1}$. Unlike the more conventional STEM diffraction setup, $\Delta q_\alpha$ will in our case be reduced with increasing convergence angle $\alpha$, while the spatial resolution is compromised as a larger sample area is illuminated. The spatial resolution is then

$$\Delta x \approx z\alpha \qquad (S19)$$

resulting in an illuminated area of $\sim 500\ nm$ at $z = 338\ \mu m$.

The contribution of the finite EELS entrance aperture size is dependent on the magnification of the projector lenses and the smallest available EELS entrance aperture size (1 mm):

$$\Delta q_\beta = \frac{d_{EELS\ aperture}}{MAG} \times \frac{k_i}{z} \qquad (S20)$$

The smallest $\Delta q_\beta$ in this work is $\sim 0.0036$ Å$^{-1}$ at $z = 338\ \mu m$ and $MAG = 27000\ \times$, limited by the stability of the projection system. The total momentum resolution is

$$\Delta q = \Delta q_\alpha + \Delta q_\beta \qquad (S21)$$

To evaluate the effect of finite $\Delta q$ on the accuracy of the plasmon peak position and peak width, we simulate the EELS spectra summed within an aperture of finite collection angle $\Delta q_\beta$ from the EELS probability density in SI Section 2b, convoluted with a Gaussian diffraction spot with size $\Delta q_\alpha$ (Fig. S5a,b). The peak position and peak width of the simulated spectra are compared with the analytical model to estimate the effect of finite $\Delta q$ on the extracted peak position ($\Delta\omega_p$) and peak width ($\Delta\Gamma$) (Fig. S5c,d) and are added to the corresponding error bars of the plots in Figs. 2 and 3.

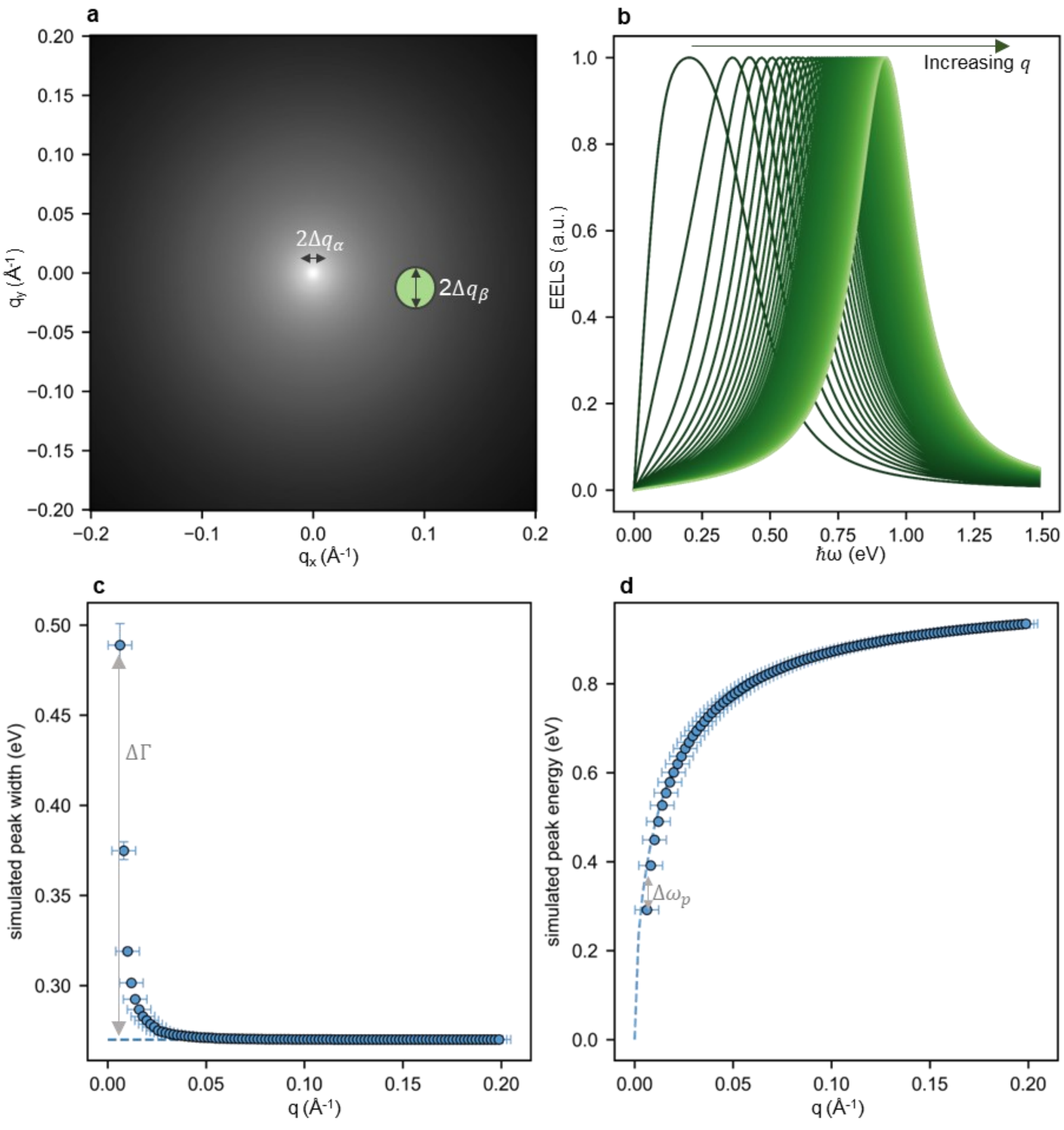

*Fig. S5 | Effect of the finite $\Delta q$ resolution on the EELS spectra. (a) The EELS spectra are summed within an aperture of finite collection angle $\Delta q_\beta$ (green disk) from the EELS probability density $\frac{d^2P}{d\Omega dE}(\omega, \boldsymbol{q})$ in Equation (S12), convoluted with a Gaussian diffraction spot with size $\Delta q_\alpha$. The resulting summed EELS spectra are presented in (b). The peak width of the spectra in (b) are presented as a scatterplot in (c), with the dispersed peak position (i.e. energy) in (d), compared with nominal values as dashed blue lines to estimate the uncertainty in plasmon peak widths and position due to the finite $\Delta q$ resolution.*

## 4. EELS Spectral Processing

### a. NLLS co-fitting

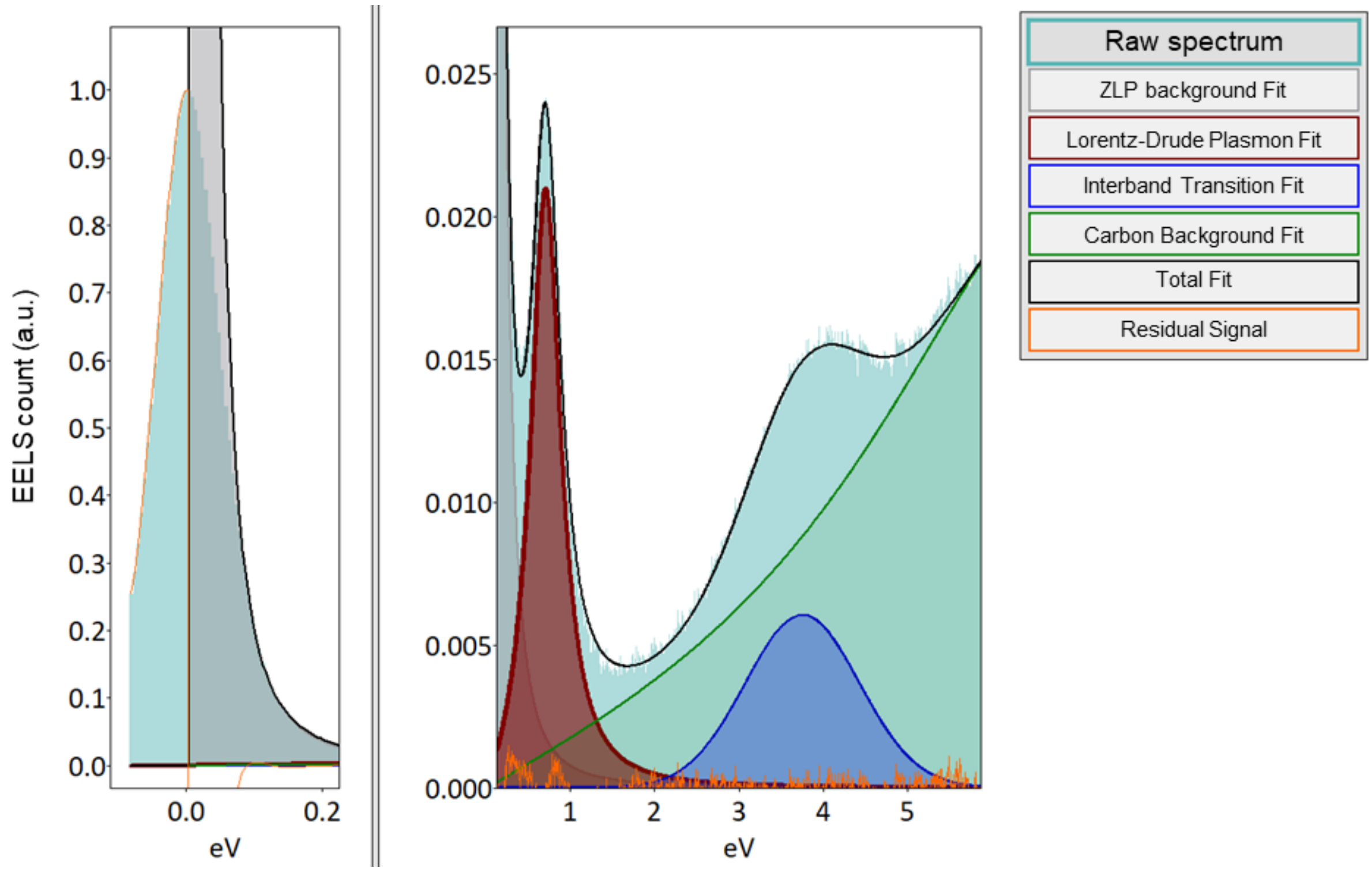

*Fig. S6 | Plasmon peak fitting via NLLS co-fitting using fixed fitting windows for all spectra. The ZLP background is fitted with a power law function as done in previous momentum-resolved EELS measurements[28]. The plasmon peak is fitted with a Lorentz-Drude function as in Equation (S16) The interband transition is fitted with a Gaussian function and the π plasmon of the carbon background (Fig. S3c) is fitted with a Lorentz-Drude function.*

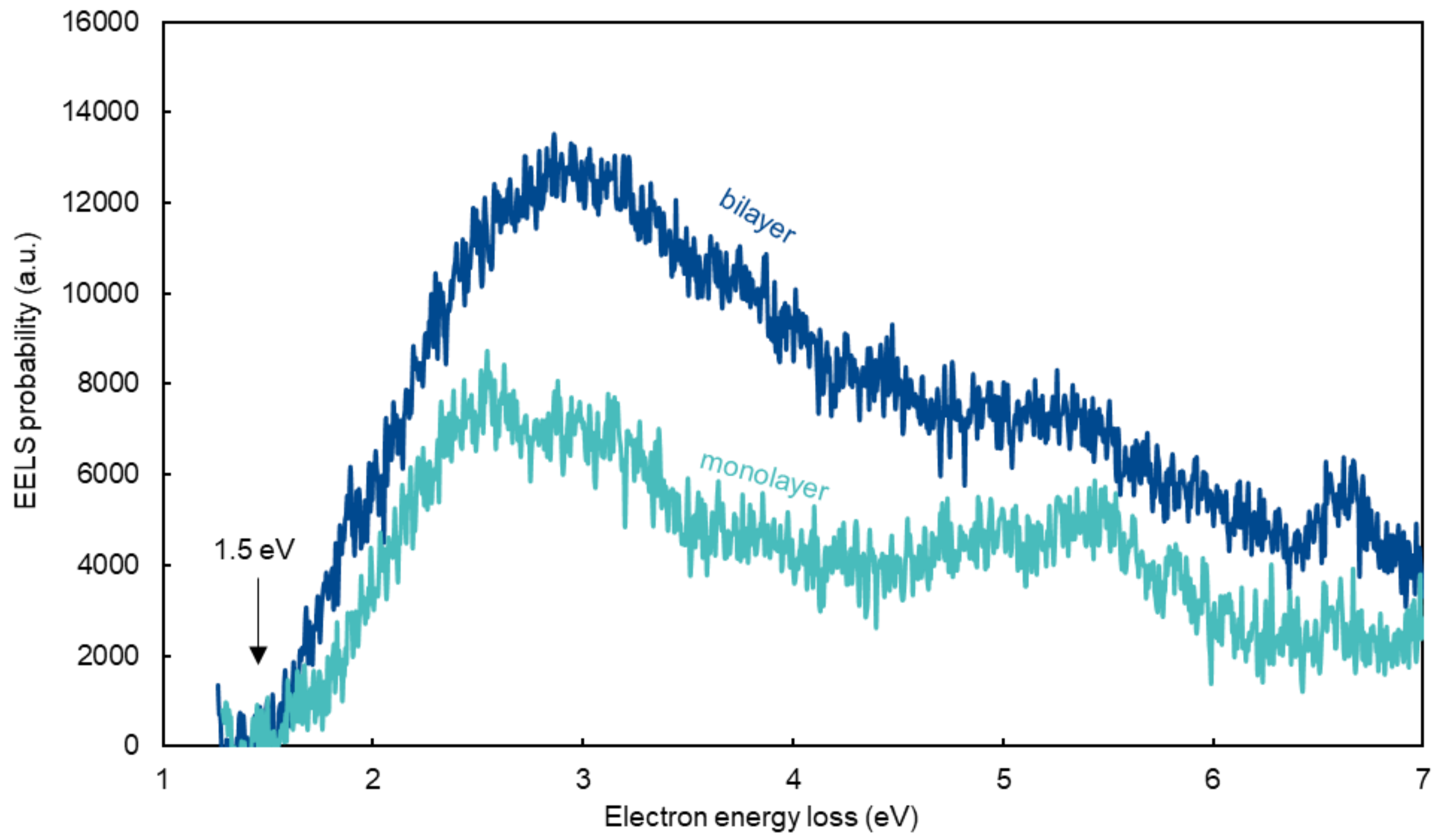

*Fig. S7 | Interband transition of monolayer (cyan) and bilayer (blue) 2H-$TaS_2$ measured at q=0. The onset is at 1.5 eV, well above the plasmon energy.*

## b. q-dependent intensity normalization

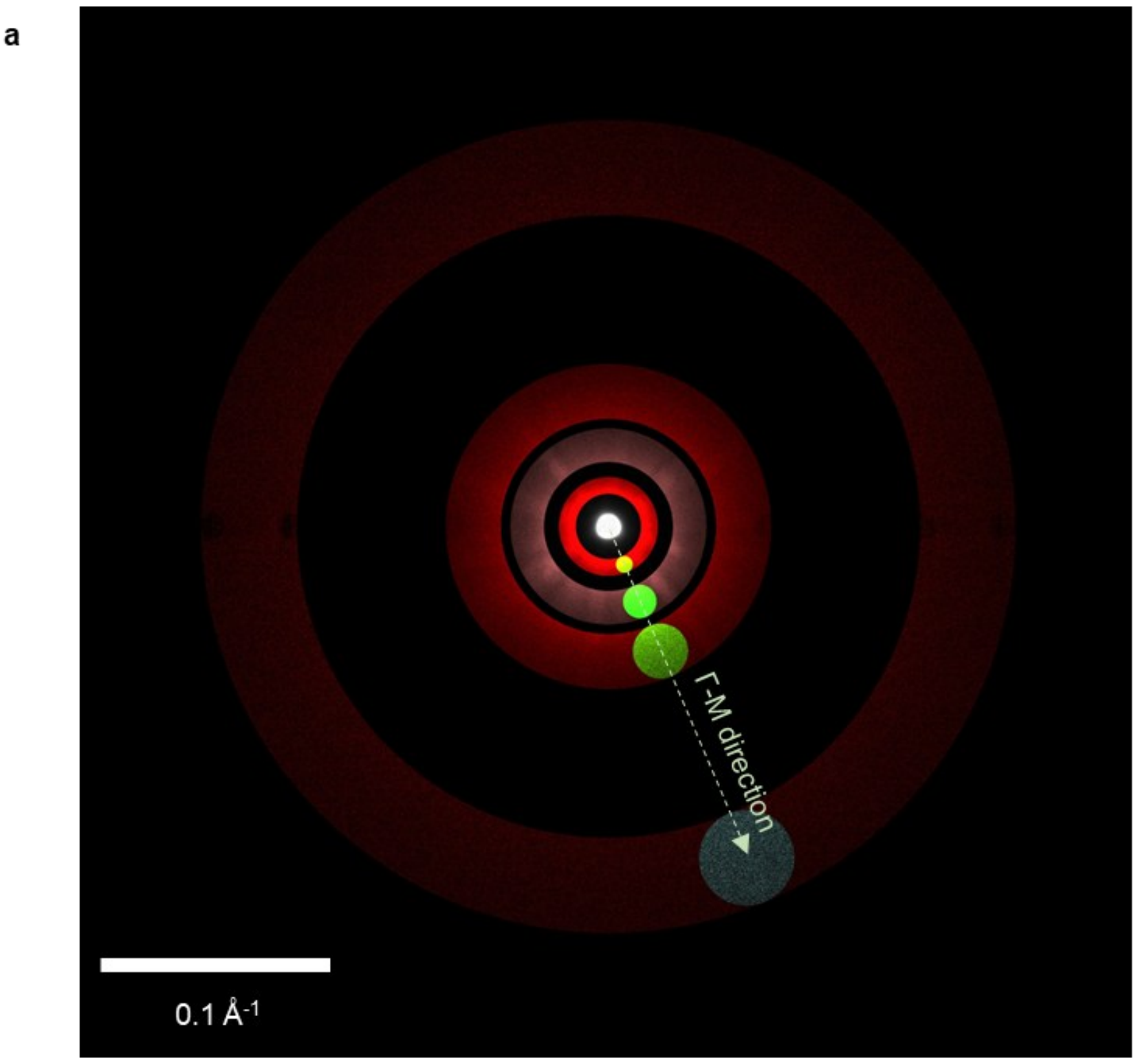

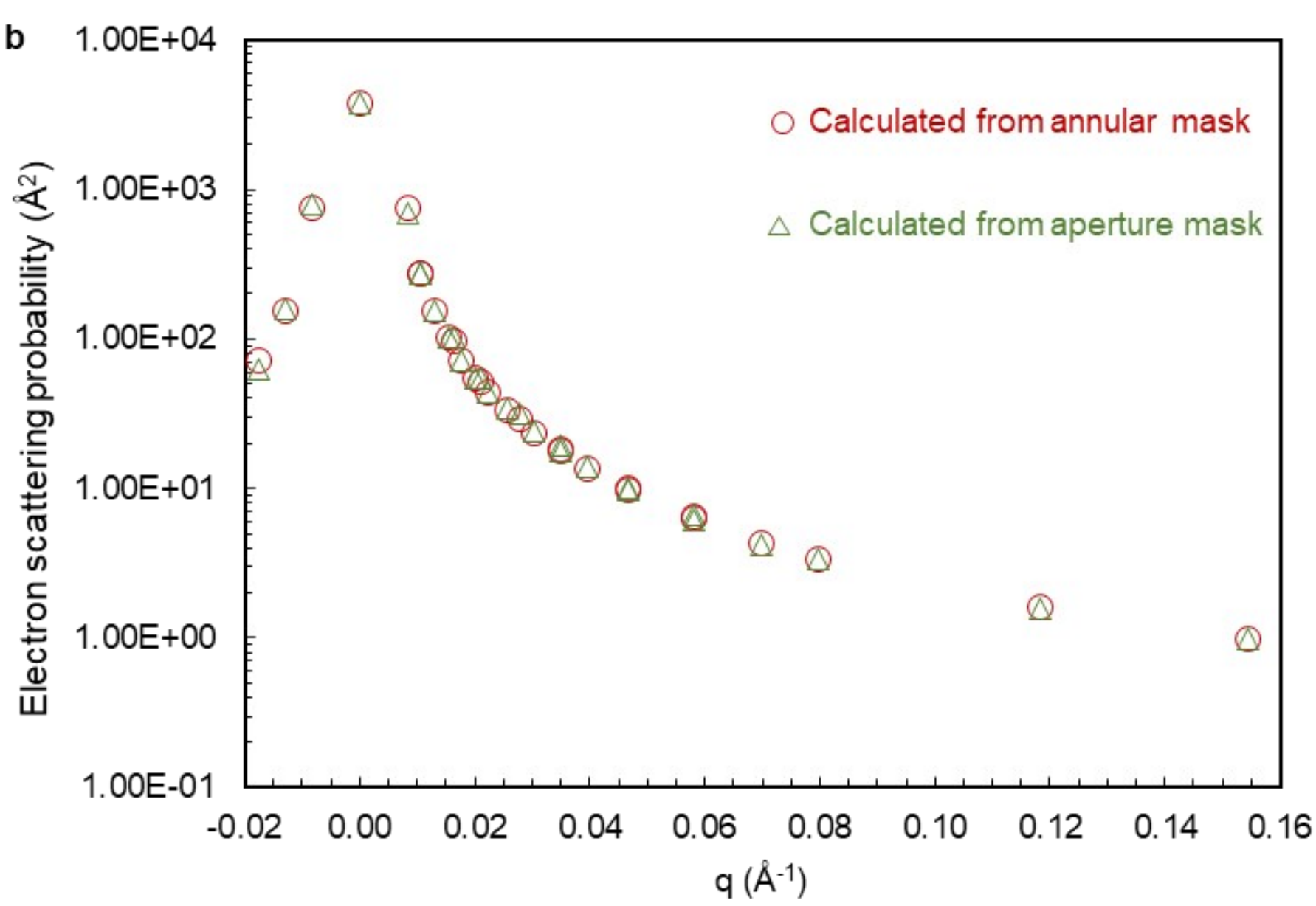

*Fig. S8 | As the EELS spectra are acquired with a few different q-resolution settings, renormalization is needed to obtain the differential loss probability* $\frac{d^2P}{d\Omega dE}(q)$*. We first normalized each spectrum with* $\int \frac{dP'(q)}{dE} dE = 1$*, then multiplied it with the q-dependent differential scattering probability* $P'(q) = \frac{dP}{d\Omega}(q)$ *with* $\int \frac{dP}{d\Omega}(q) d\Omega = 1$*. We calculated* $P'(q)$ *from an image in q-space taken with a CMOS camera. The calculation is done using both an annular mask* $d\Omega = 2\pi q dq$ *and an aperture mask with* $d\Omega$ *matching the size of the EELS entrance aperture (a). Panel (b) shows* $P'(q)$ *with consistent results between two types of masks, demonstrating the robustness of the normalization.*

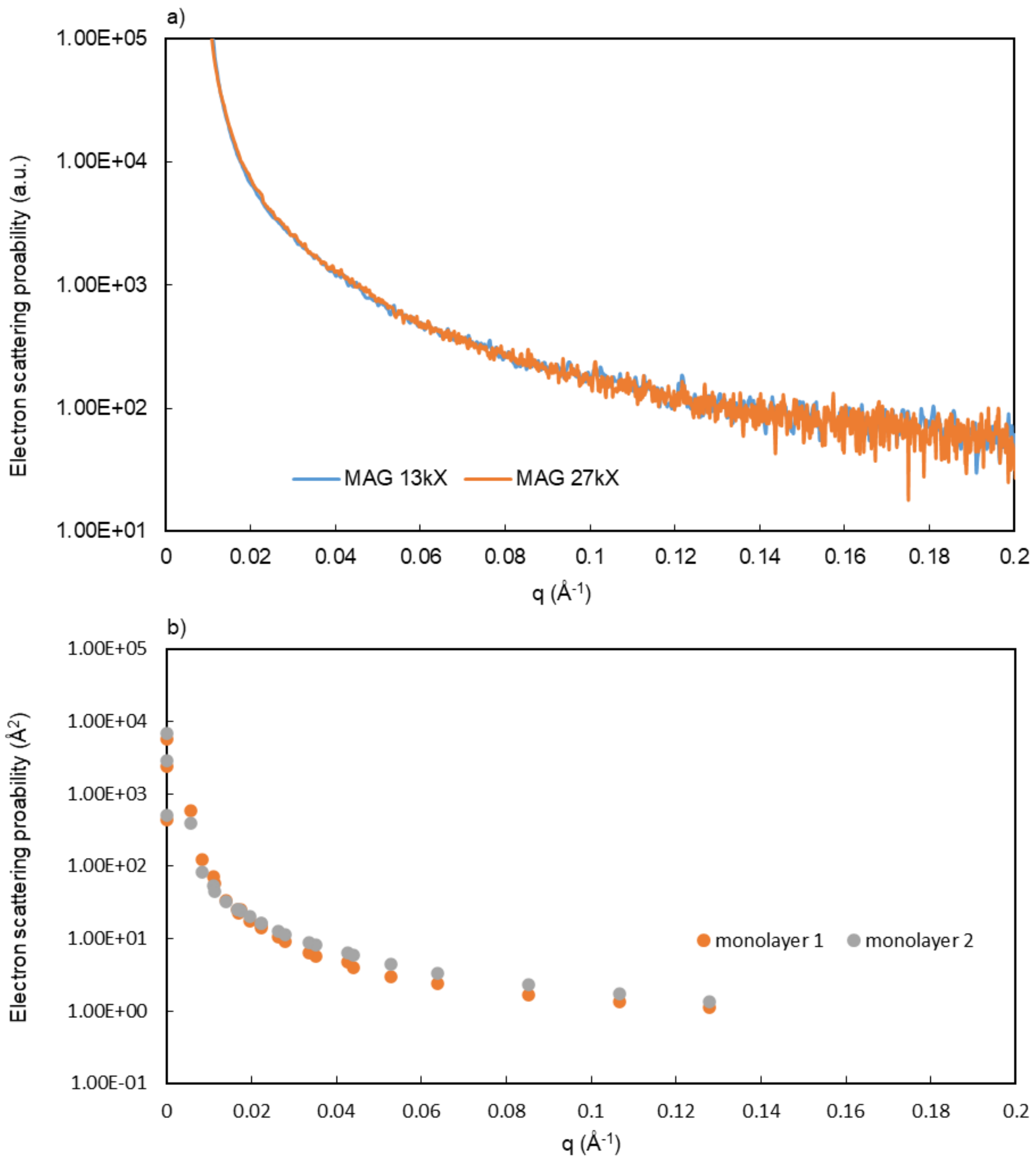

*Fig. 9 | The measured electron scattering probability* $\frac{dP}{d\Omega}(q)$ *is consistent across a) different projector magnifications used in the experiment, and b) different samples, showing the robustness of the normalization.*

## 5. Plasmon dispersion fitting and interband screening extraction

The nonlocal interband screening in this work can be characterized with an effective screening dielectric function $\varepsilon_{inter}(q,\omega)$ that screens the interaction between free charge carriers forming plasmons. As the plasmon energy is below the interband transition onset, and much lower than the peak of the interband transition at $\sim 3eV$ observed in the EELS spectra (Fig. S7), we can safely assume a static screening $\varepsilon_{inter}(q)$ without frequency-dependence. The effect of interband screening $\varepsilon_{inter}(q)$ on plasmon dispersion relation is shown analytically in Equation (S6). In the low$-q$ limit ($qd \ll 1$), $\varepsilon_{inter}(q)$ can be represented analytically with the Keldysh model where the 2D material is described as a slab of constant dielectric function[29]. For an ideal 2D monolayer $d \to 0$, $\varepsilon_{inter}(q)$ can be expressed as[30]

$$\varepsilon_{inter}(q) = 1 + \rho_0 q \tag{S22}$$

Where $\rho_0 = \frac{d\varepsilon_{bulk}}{2}$ is the characteristic screening length. More accurate *ab initio* calculations show the deviation from the Keldysh model of $\varepsilon_{inter}(q)$ at large $q$ due to finite thickness of the real 2D material [31]. Experimentally, the deviation has been shown indirectly[32] where Keldysh model underestimates the influence of environmental screening on binding energy of excitons in $WSe_2$ monolayers.

We now apply two fitting procedures: first for a fixed, then for a fitted screening length. In the first fitting procedure, Equation (S6) is rewritten with the term $\varepsilon_{inter}(q)$ expressed in Equation (S22) with fixed effective screening lengths $\rho_{0,ML} = 28.4$ Å for monolayers and $\rho_{0,BL} = 56.8$ Å for bilayers, from da Jornada et al.[9] Orthogonal distance regression is applied to the measured data with the Drude weight as fitting parameter. The results are presented as striped lines in Fig. 2b. From the extracted Drude weights, the corresponding interband screening is calculated and plotted in Fig. 2c. The second fitting procedure treats $\rho_0$ as a fitting parameter on the low$-q$ regime ($q < 0.05$ Å$^{-1}$), for which we can assume a linear relationship between $q$ and the interband screening $\varepsilon_{inter}$. The results are also plotted in Fig. S10, showing a robust extraction of the interband screening within the error bars.

We performed the fitting via the Orthogonal Distance Regression method [a] that takes into consideration the error bars in both the x- and y- axes. The y-axis error bars for the plasmon frequency include uncertainty from the NLLS peak fitting, the width of the ZLP and from the effect of finite $\Delta q$ resolution. The $q$ $-$dependence of the EELS probability density $\frac{d^2P}{d\Omega dE}$ results in an underestimation of the plasmon energy and broadening of the plasmon peak integrated in a finite aperture with radius $\Delta q_\beta$, especially in the low $q$ regime, as numerically demonstrated in SI Section 3b.

The fitting parameters for the dispersion relation routines are presented in Table 1 below.

Here, we attempt to extract the nonlocal screening in monolayers and bilayers of 2H-$TaS_2$ by comparing the measured screened dispersion relation with the $\sim\sqrt{q}$ dispersion relation of an ideal 2D electron gas from Equation (S6) using the fitted values of $D$, presented in Fig. S10c, d:

$$\varepsilon_{inter}(\boldsymbol{q}) = \frac{Dq}{2\pi\varepsilon_0\omega_p^2(q)} \tag{S23}$$

[a] https://docs.scipy.org/doc/scipy/reference/odr.html

The Drude weight $\mathcal{D} = \pi e^2 \left[\frac{n}{m}\right](q)_{eff}$ is assumed to be independent of $q$ as the wave vector range of interest only covers a small part of the Brillouin zone as indicated in the diffraction pattern in Fig. 2a inset.

*Table S1. Fitting parameters for the dispersion relation routines presented in Fig. S10.*

| | | Screening length $\rho_0(\text{Å}^{-1})$ | Drude weight $\mathcal{D}(\times 10^{10}\ \Omega^{-1}s^{-1})$ |
|---|---|---|---|
| Routine 1: **Fixed** screening length | Monolayer | 28.4 | 31.2 ± 0.8 |
| | Bilayer | 56.8 | 67.8 ± 1.2 |
| Routine 2: **Fitted** screening length | Monolayer | 30.8 ± 9.7 | 33.8 ± 4.8 |
| | Bilayer | 58.3 ± 5.4 | 63.0 ± 3.8 |

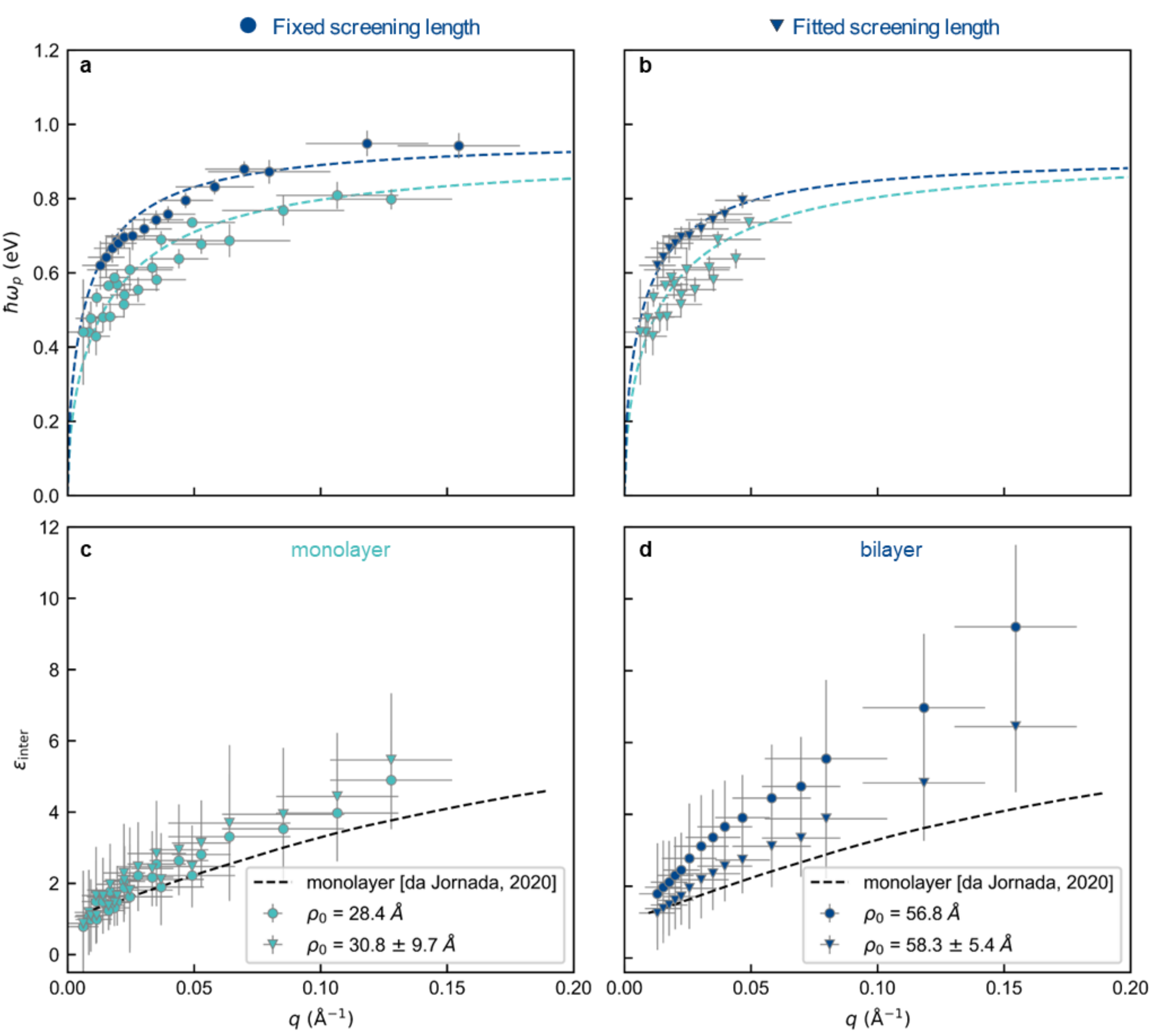

*Fig. S10 | Dispersion relation fitting (a,b) and interband screening extraction (c,d) with two different fitting routines.*

## 6. EELS prefactor extraction

### c. Oversampling in q-space

As we obtained EELS spectra at varying $\Delta q$ resolution to account for low signal in the high-$q$ regime, we binned the spectra in the $q$ −space. In the main text, selected data points are presented for clarity. In this section, we provide the full dataset for the bilayer sample, demonstrating the consistency of the results under different acquisition conditions, as shown in Fig. S11. The data series were collected from low $\Delta q$ to high $\Delta q$, and the overlap of different $\Delta q$ data series highlights the robustness of the results against electron dose.

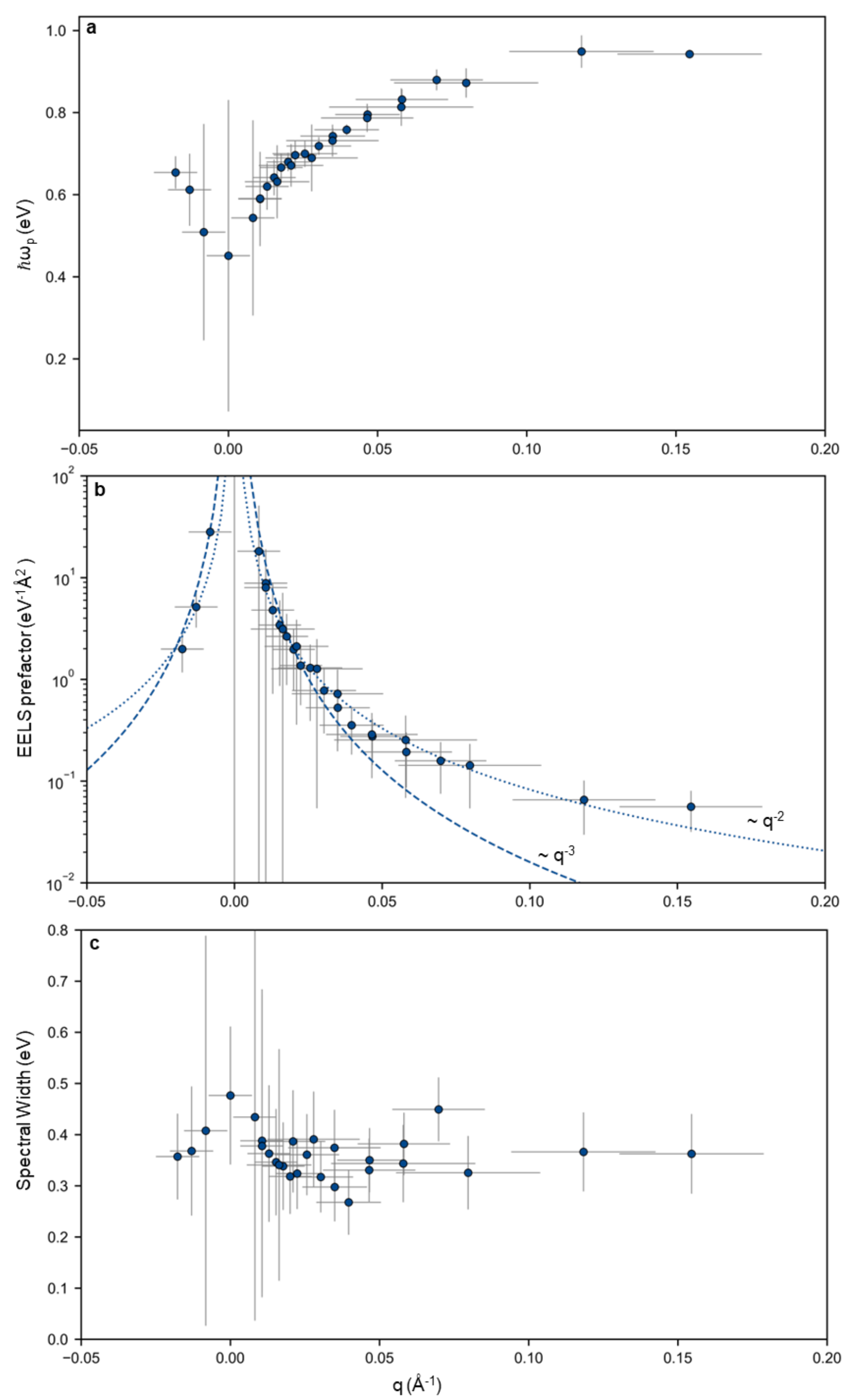

*Fig. S11 | (a) Extracted peak position, (b) EELS prefactor, and (c) peak width from the oversampled dataset with overlapping datapoints under different acquisition conditions.*

## d. Effect of intensity re-normalization

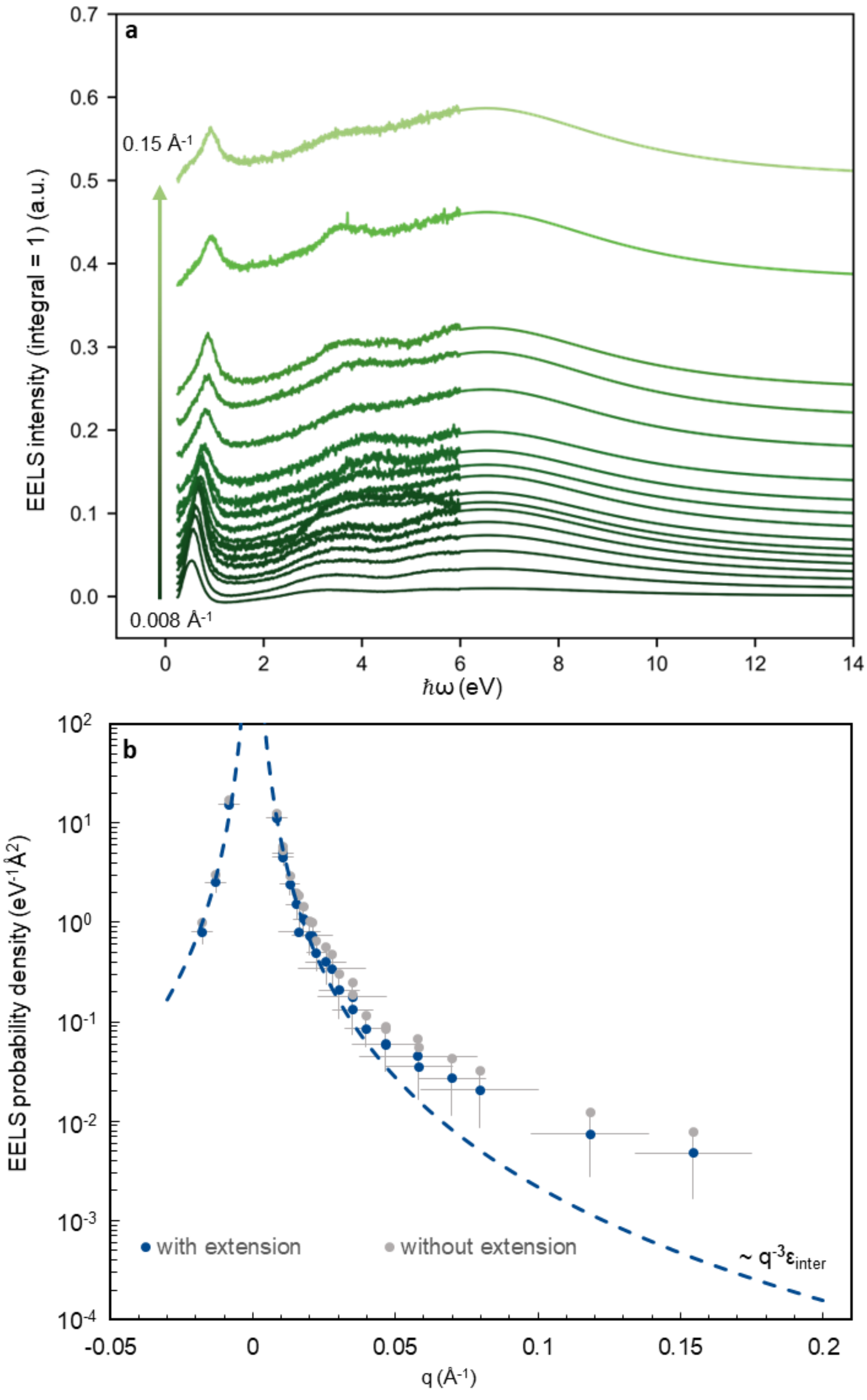

*Fig. S12 | Extrapolating the carbon plasmon peak prevents an underestimation of the normalization with the integral of the EELS spectra. The error bars in the spectral weight are estimated from the difference between the cases with and without the extension of the carbon plasmon peak. a) EELS spectra after the extension with the fitted carbon plasmon peak. b) Peak intensity of the EELS probability density* $\frac{d^2P}{d\Omega dE}$ *with and without the extension.*

## 7. Reproducibility of monolayer results

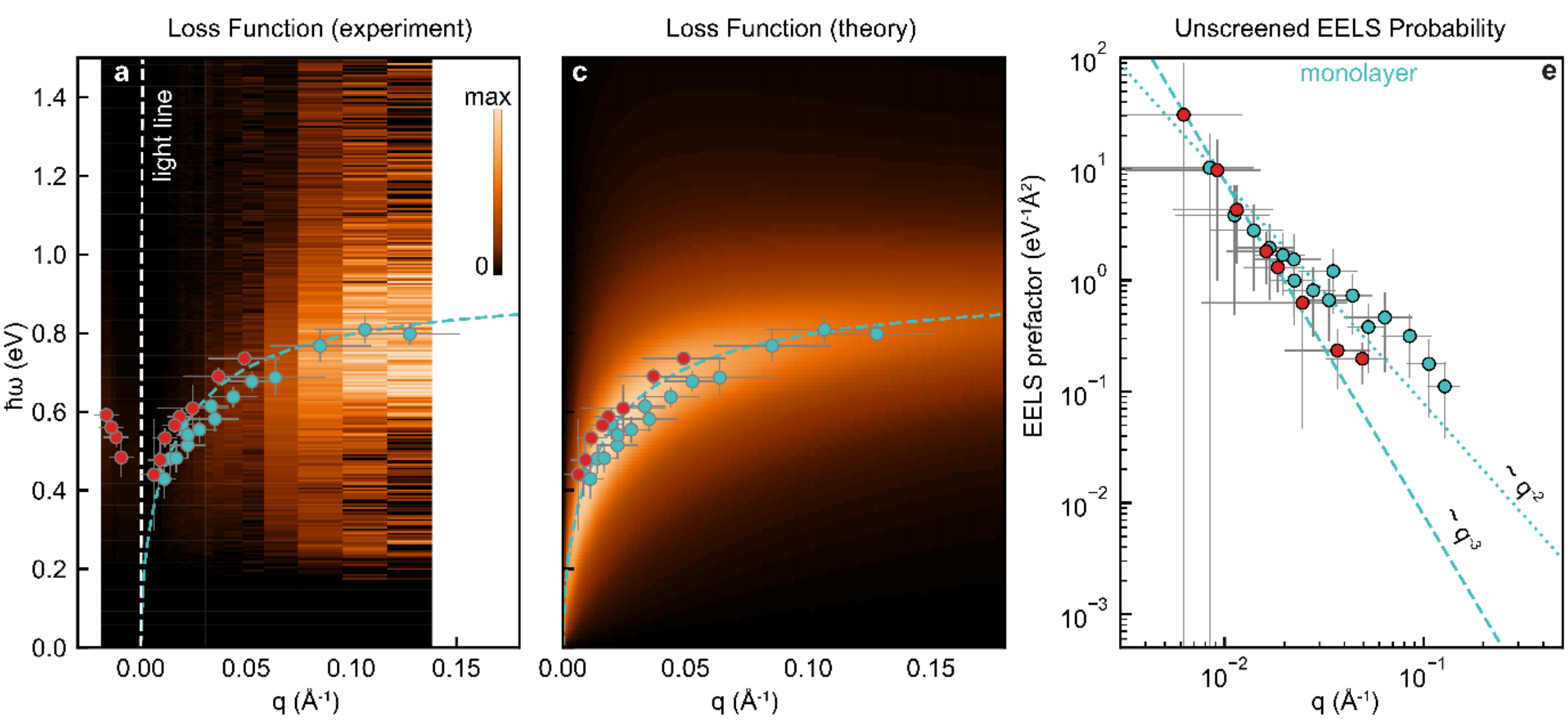

*Fig. S13 | Reproducibility of momentum-resolved EELS in $TaS_2$ monolayers. Comparison between the two monolayer datasets of 2 samples in red and cyan, demonstrating reproducibility of the momentum-resolved EELS results. a) Experimental loss function extracted from the measured EELS intensity after removing the EELS prefactor (~$q^{-3}$). b) Simulated loss function using the Lorentz-Drude model with fitted Drude weights and interband screening from both datasets. c) Extracted EELS prefactor from the fitted Lorentz-Drude amplitudes after removing the interband screening factor. The crossover from ~$q^{-3}$ to ~$q^{-2}$ scaling occurs at similar wave vectors in both datasets*

## 8. Temperature-dependent plasmon peak width for 2H-TaS2

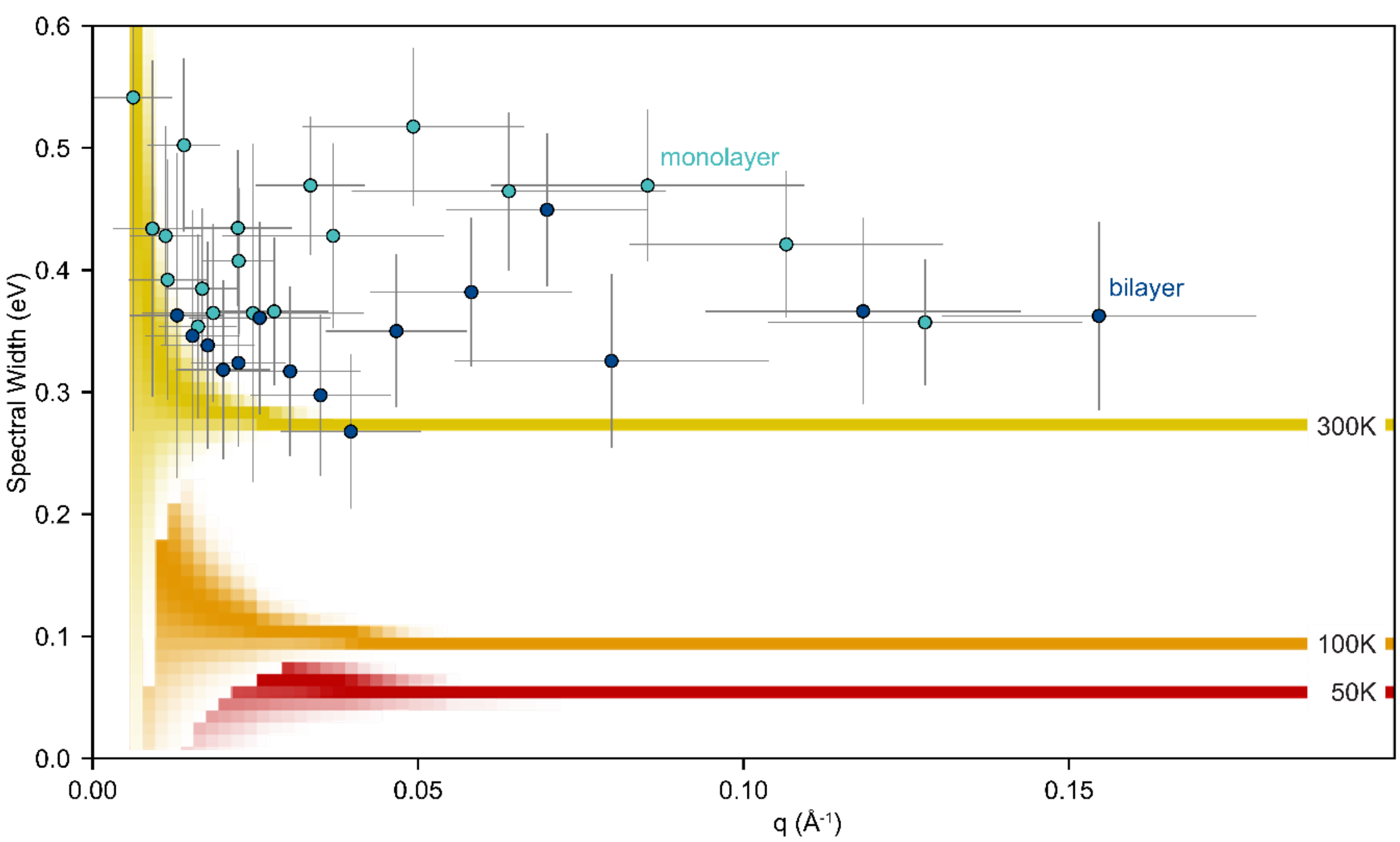

*Fig. S14 | Measured q-dependent plasmon peak widths of 2H-$TaS_2$ monolayers and bilayers. The coloured plots show simulated peak width broadening due to the finite size of the EELS entrance aperture (Fig. S5) at varying temperatures. The calculated peak widths for $\Gamma_{q\to\infty}$ are 0.27 eV, 0.09 eV and 0.045 eV for T=300 K, 100 K and 50 K, respectively, from calculated relaxation times by Hinsche et al.[33] for $TaS_2$ monolayers. Instrumental broadening is removed through an empirical estimation from the FWHM of the ZLP shown in Eq (5) of Bosman et al.[34].*

## 9. Electron-beam-enhanced carbon signal

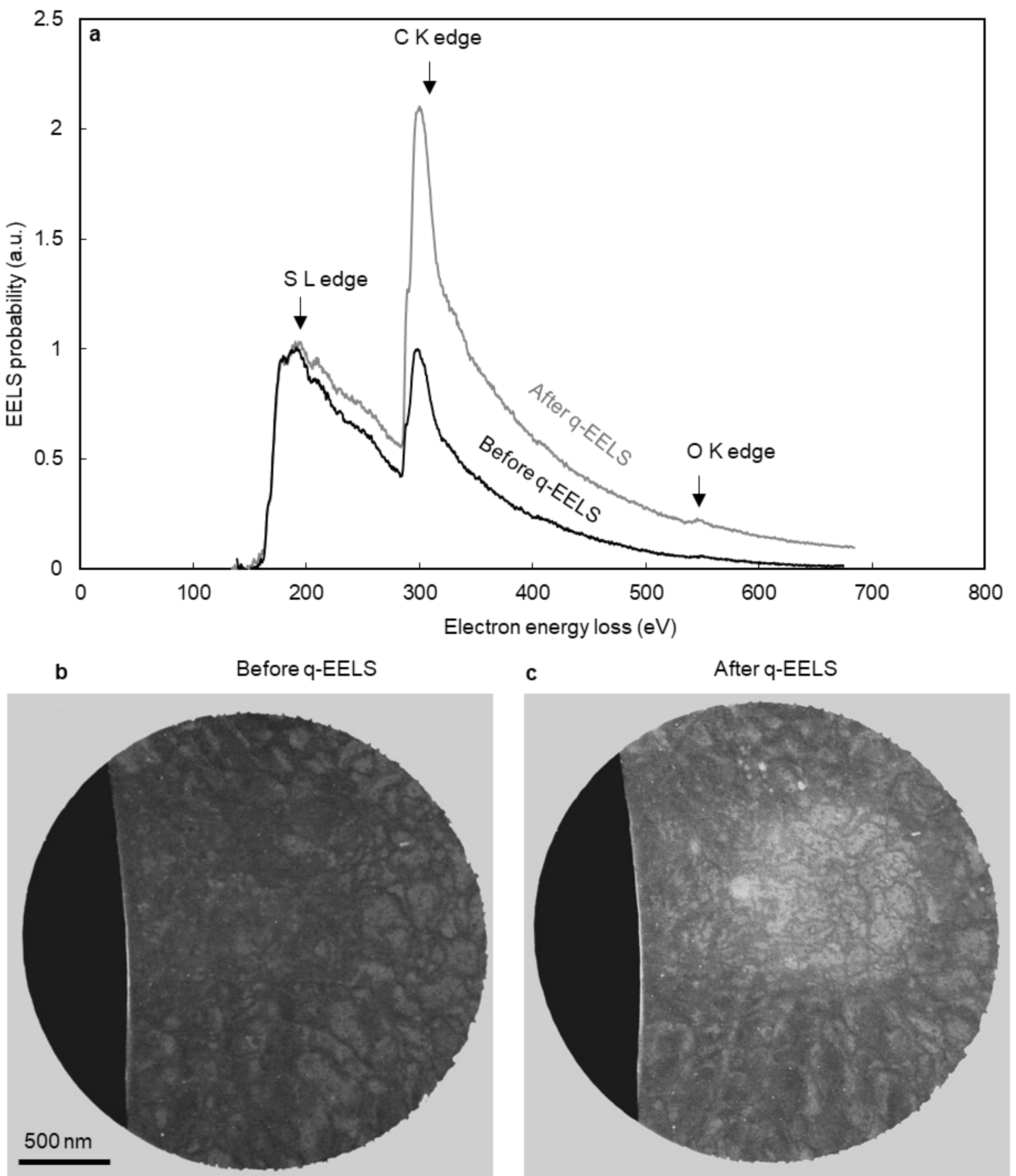

*Fig. S15 | a) Core-loss EELS and e-beam induced carbon signal after 4 hours of q-EELS measurements, resulting in an increase of C K and O K edges from the solvent used to disperse the exfoliated flakes (propylene carbonate). b) and c) Low angle, annular dark field (LAADF) images of the bilayer sample before and after q-EELS experiment, respectively. LAADF images emphasize light-element scattering in comparison with the usual high-angle, annular dark field (HAADF) STEM images.*

## 10. Dose effect on the EELS spectrum

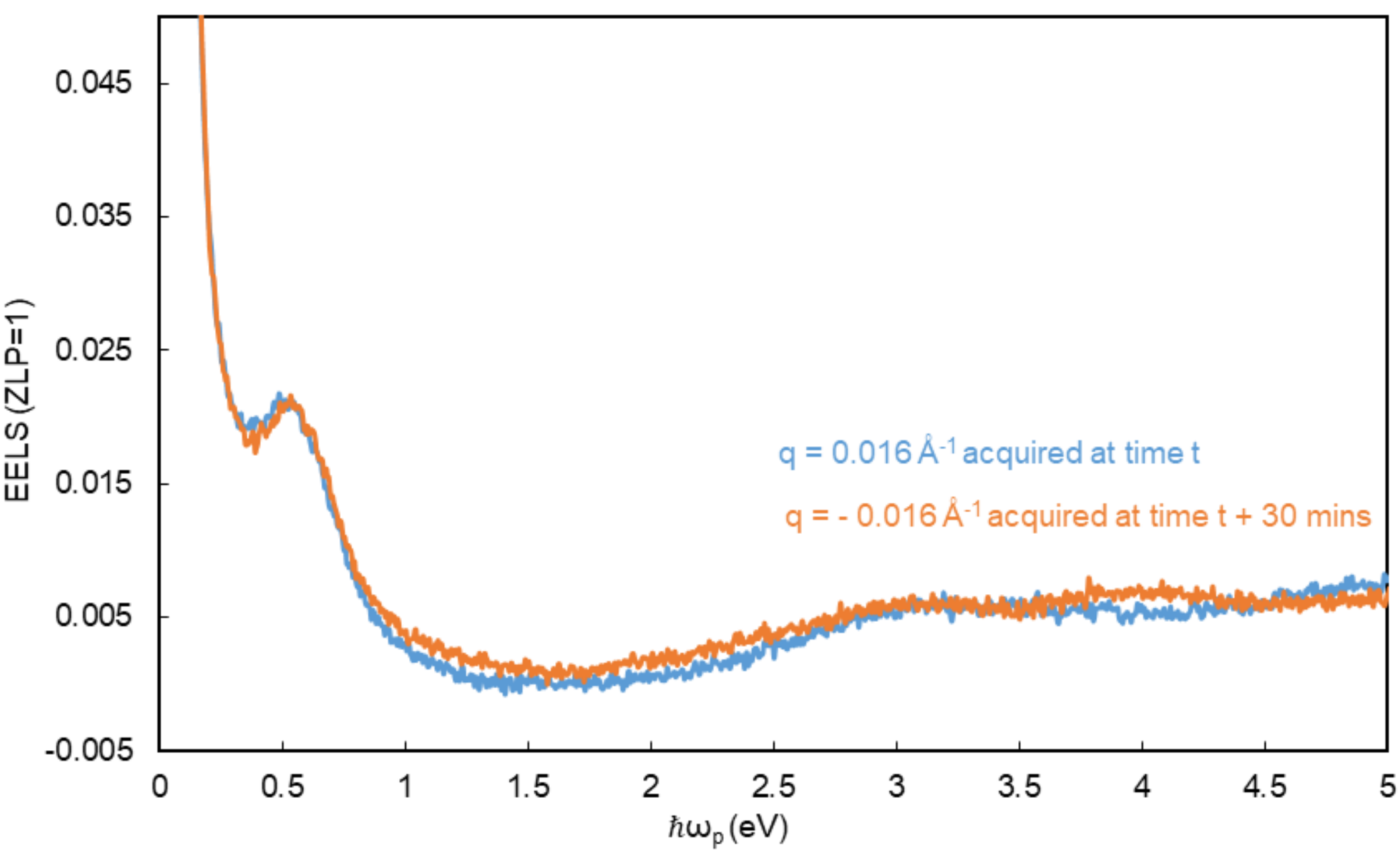

*Fig. S16 | Dose effect. EELS Spectra of the monolayer acquired at* $q = 0.16$ Å$^{-1}$ and $q = -0.16$ Å$^{-1}$, 30 minutes apart.